\documentclass{aastex701}

\usepackage{amsmath}
\usepackage{graphicx}
\usepackage{xcolor}
\usepackage{siunitx}
\DeclareRobustCommand{\sampleWFD}{WFD-only}
\DeclareRobustCommand{\sampleTWWFD}{TW+WFD}
\DeclareRobustCommand{\sampleTWTNS}{TW+TNS}
\DeclareRobustCommand{\sampleCombined}{TW+WFD+TNS}
\journalinfo{}
\begin{document}

\title{Design and Forecasts for a Rubin Twilight Program for Nearby Type Ia Supernovae}

\author[0009-0007-3185-7030]{Xianzhe TZ Tang}
\email{tztang@bu.edu}
\affiliation{Boston University Department of Astronomy, 685 Commonwealth Avenue, Boston, MA 02215, USA}
\affiliation{Institute for Astrophysical Research, 685 Commonwealth Avenue, Boston, MA 02215, USA}

\author[0000-0001-5201-8374]{Dillon Brout}
\email{dbrout@bu.edu}
\affiliation{Boston University Department of Astronomy, 685 Commonwealth Avenue, Boston, MA 02215, USA}
\affiliation{Institute for Astrophysical Research, 685 Commonwealth Avenue, Boston, MA 02215, USA}
\affiliation{Boston University Department of Physics, 725 Commonwealth Avenue, Boston, MA 02215, USA}

\author[0000-0003-0347-1724]{Christopher W. Stubbs}
\email{stubbs@g.harvard.edu}
\affiliation{Department of Physics, Harvard University, 17 Oxford Street, Cambridge, MA 02138, USA}
\affiliation{Department of Astronomy, Harvard University, 60 Garden Street, Cambridge, MA 02138, USA}

\author[0000-0002-3205-2484]{Elana K. Urbach}
\email{eurbach@g.harvard.edu}
\affiliation{Department of Physics, Harvard University, 17 Oxford Street, Cambridge, MA 02138, USA}

\author[0000-0002-6825-5283]{J. Bryce Kalmbach}
\email{brycek@slac.stanford.edu}
\affiliation{SLAC National Accelerator Laboratory, 2575 Sand Hill Road, Menlo Park, CA 94025, USA}
\affiliation{Kavli Institute for Particle Astrophysics and Cosmology (KIPAC), Stanford University, Stanford, CA 94305, USA}
\affiliation{DiRAC Institute, University of Washington, Seattle, WA 98195, USA}

\author[0009-0009-1689-4874]{Greg Aldering}
\email{galdering@lbl.gov}
\affiliation{Lawrence Berkeley National Laboratory, 1 Cyclotron Road, Berkeley, CA 94720, USA}

\author[0000-0002-7234-844X]{Bastien Carreres}
\email{bastien.carreres@duke.edu}
\affiliation{Duke University, Durham, NC 27708, USA}

\author[0000-0001-9553-4723]{Chris Frohmaier}
\email{chris.frohmaier@port.ac.uk}
\affiliation{Institute of Cosmology and Gravitation, University of Portsmouth, Burnaby Road, Dennis Sciama Building, Portsmouth PO1 3FX, UK}

\author[0000-0002-9436-8871]{Ayan Mitra}
\email{ayan@illinois.edu}
\affiliation{National Center for Supercomputing Applications, University of Illinois at Urbana-Champaign, 1205 W. Clark St., MC-257 Room 1008, Urbana, IL 61801, USA}

\author[0000-0001-8596-4746]{Erik R. Peterson}
\email{erikpete@umich.edu}
\affiliation{Department of Physics, University of Michigan, Ann Arbor, MI 48109, USA}
\affiliation{Society of Fellows, University of Michigan, Ann Arbor, MI 48109, USA}

\author[0000-0002-3500-6635]{Corentin Ravoux}
\email{corentin.ravoux@clermont.in2p3.fr}
\affiliation{Universit\'e Clermont-Auvergne, CNRS, LPCA, 63000 Clermont-Ferrand, France}

\author{the LSST Dark Energy Science Collaboration}
\email{}
\noaffiliation
\makeatletter
\expandafter\gdef\csname affil@\romannumeral\the\allauthorcnt\endcsname{}
\makeatother

\date{\today}

\begin{abstract}

Nearby Type~Ia supernovae (SNe~Ia) are important for supernova cosmology, but saturation in 30\,s exposures, gaps in light-curve sampling, and the footprint of Vera C. Rubin Observatory's baseline Wide-Fast-Deep (WFD) survey limit the nearby sample.
To improve the low-$z$ sample, we propose a twilight follow-up program using Rubin Observatory for 1--5\,s \emph{griz} exposures at solar altitudes $-12^\circ\!\le\!\mathrm{Alt}_\odot\!\le\!-9^\circ$.
Using simulations of one year of WFD observations, we find that twilight follow-up adds 230 quality-selected SNe~Ia from an external sample drawn primarily from the Transient Name Server (TNS). At $z<0.08$, the usable sample increases from 368 to 643 SNe~Ia, including 50 WFD events recovered by refitting their light curves with twilight photometry. Combining these twilight observations with high-redshift WFD and Deep Drilling Field samples increases the statistical-only dark-energy $(w_0,w_a)$ figure of merit from 49.3 to 58.3 (18\%) and reduces $\sigma(w_a)$ by 10\%. For measurements of structure growth at $0.02\le z<0.06$, the fractional uncertainty on $f\sigma_8$ decreases from 0.783 to 0.305. Restricting the program to $-12^\circ\le\mathrm{Alt}_\odot\le-11^\circ$ reduces the dark-energy improvement to 7.2\%. These short exposures could also provide unsaturated photometry of potential distance-ladder calibrators at $z<0.02$.
\end{abstract}

\keywords{\uat{Type Ia supernovae}{1728} --- \uat{Sky surveys}{1464} --- \uat{Observational cosmology}{1146} --- \uat{Photometry}{1234} --- \uat{Dark energy}{351}}

\section{Introduction}
Type~Ia supernovae (SNe~Ia) measure relative luminosity distances, while baryon acoustic oscillations (BAO) measure distances and expansion rates in units of the sound horizon. Recent analyses combining these probes report evidence for time-varying dark energy whose significance depends on the SN sample: DESI~DR2 combinations with three current SN compilations span about $2.8\sigma$ to $4.2\sigma$ \citep{brout2022pantheonplus,union,des5yrcosmology,desi_dr2_bao_2503}. A well-measured nearby sample establishes the low-redshift end of the relative SN Hubble diagram against which the higher-redshift distances are compared. Analyses that change the calibration, treatment, or composition of the nearby sample lead to corresponding shifts in the inferred evidence for time-varying dark energy \citep{huang_2025_lowz_sys,zkh_2025gyn,vincenzi_response_25}. Cross-probe consistency tests further motivate improving the nearby sample \citep{tz_uniting}. The Zwicky Transient Facility (ZTF; \citealt{Bellm2019_ZTF,Fremling2020_ZTFBTS_I}) and the Young Supernova Experiment (YSE; \citealt{Jones2021_YSE}) provide nearby discovery streams and multi-band light curves. Programs such as the Dark Energy Bedrock All-Sky Supernova program (DEBASS; \citealt{lowz_shared_2508_10878,acevedo2025dark}) are being developed to obtain larger and more homogeneous nearby samples. The Type Ia supernova Trove from ATLAS in the Nearby universe (TITAN) is also preparing a large low-redshift SN~Ia sample with recalibrated photometry for cosmology \citep{Marlin2026_TITAN}.

We use the term ``nearby'' to refer to the $z<0.1$ SNe~Ia targeted by the Twilight (TW) program proposed here; most of the simulated contribution lies at $z<0.08$. These SNe also provide distances for peculiar-velocity (PV) studies of structure growth. Calibration, selection, redshift corrections, and light-curve modeling affect these applications in different ways. Rubin Observatory's Legacy Survey of Space and Time (LSST) is a ten-year survey in the six $ugrizy$ bands \citep{ivezic2019lsst}. Rubin observations on a common photometric system would provide a larger, more homogeneous nearby sample and could contribute to future training of SN~Ia light-curve models \citep{Jones2019_Foundation,Kenworthy2021_SALT3}.

The relevant redshift range depends on the application. Distance-ladder measurements of $H_0$ require SNe in hosts with independently determined distances to calibrate the SN absolute magnitude and a separate Hubble-flow sample to transfer that calibration to the expansion rate. Events at $z<0.02$ and near maximum brightness are among those most likely to saturate in 30\,s Rubin visits, and the resulting loss of high-SNR multiband measurements can compromise light-curve standardization. Short twilight exposures reduce that risk for potential calibrator-host events, although we do not forecast how many such hosts will occur. In contrast to distance-ladder analyses, dark-energy analyses use the statistical weight and calibration quality of the full low-redshift Hubble diagram. PV analyses depend on the number, sky distribution, distance precision, and velocity modeling of well-measured SNe at $z\lesssim0.1$ \citep{pv_with_snia_2020,howlett2017_pv_surveys,carreres2023_ztf_growth,fs8_intrinsic_scatter_2505_13290,fs8_forecast_2507_00157}. Throughout this paper, ``low-redshift anchor'' denotes the nearby ($z<0.1$) end of the Hubble diagram. Here we simulate a proposed Rubin TW program and estimate its effect on SN-only dark-energy and PV Fisher forecasts. Potential applications to distance-ladder calibration and SN light-curve model training are discussed in Section~\ref{sec:discussion} and excluded from these forecasts.

Rubin's baseline Wide-Fast-Deep (WFD) survey will discover and observe many SNe, but 30\,s $gri$ measurements of the brightest nearby events can saturate near maximum. The planned rolling cadence shifts the observing effort among WFD regions between seasons, while the Deep Drilling Fields (DDFs) repeatedly observe a few small fields at higher cadence. Both WFD and DDF use 30\,s visits in $gri$ in the current baseline \citep{RubinKeyNumbers2026,RubinSurveyStrategyV5_2025}. A very nearby SN~Ia near maximum can exceed the detector saturation threshold at phases that constrain light-curve color and shape. Appendix~\ref{sec:sat_validation} presents the DP1 saturation analysis and its comparison with the simulation. A saturated epoch removes or degrades a band-specific measurement, and the multi-day WFD cadence may not provide an unsaturated replacement at a comparable phase. WFD has a fiducial main survey area of $18{,}000\,\mathrm{deg}^2$ \citep{RubinKeyNumbers2026}, DDFs provide high cadence over small areas, and regions outside WFD are not guaranteed WFD-like cadence even where mini-surveys provide some coverage \citep{ivezic2019lsst,JonesPSTN051,bianco2022rubin,RubinMiniSurveys2026}. The resulting light-curve gaps stem from peak saturation in the brightest events and sparse coverage outside the WFD footprint.

TW pointings with shorter exposure times can recover unsaturated peak-phase photometry for WFD events and follow visible nearby events outside the WFD footprint. Short-exposure twilight surveys have been proposed for near-Sun Solar System discovery \citep{near_sun_twilight_wp_2018}. Here we develop follow-up of reported nearby SNe~Ia, with exposure times set by a saturation check and visits coordinated with the WFD cadence. We adopt 1--5\,s exposures in the darker part of nautical twilight, $-12^\circ\le\mathrm{Alt}_\odot\le-9^\circ$, and limit filter changes because 120\,s is the documented maximum time for a filter swap \citep{RubinDMTN065}. Setting the final solar-altitude limits will require on-sky tests of depth, calibration, and scheduling. During Rubin operations, candidate TW triggers would be selected from reports submitted to the Transient Name Server (TNS) and from Rubin alerts. Rubin alerts add nearby WFD events whose standard 30\,s visits saturate near peak or whose scheduled cadence provides inadequate peak-phase coverage; the TW program provides short-exposure \emph{griz} follow-up within Rubin's visibility constraints.
For this study, we restrict TW visits to \emph{griz} (whereas the DP1 saturation check in Appendix~\ref{sec:sat_validation} uses the available \emph{gri} commissioning data), though the scheduler supports a configurable filter set; in future operations, matching the filter of adjoining WFD or DDF visits throughout each twilight window would avoid 120\,s intra-window swaps. All forecasts assume reliable classifications and host redshifts for selected candidates.

We evaluate this program with a scheduler developed for this work and applied to realistic simulations. The scheduler applies altitude, Moon, depth, cadence, and saturation requirements and writes the time, sky position, filter, exposure time, and observing conditions for each planned visit. We analyze the light curves with and without the proposed TW program and use the resulting distance uncertainties in SN-only dark-energy and PV forecasts. Section~\ref{sec:methods} describes the scheduler, simulations, and forecast framework; Section~\ref{sec:results} presents the one-year results; and Section~\ref{sec:discussion} examines operational constraints and broader science applications.

\section{Methodology}
\label{sec:methods}

The proposed TW program follows nearby SNe absent from the baseline WFD realization and adds short exposures to low-$z$ WFD light curves to reduce saturation near peak.

\subsection{Scheduler and Visit Planning}
\label{sec:scheduler-overview}
\begin{figure}

\centering
\includegraphics[width=\linewidth]{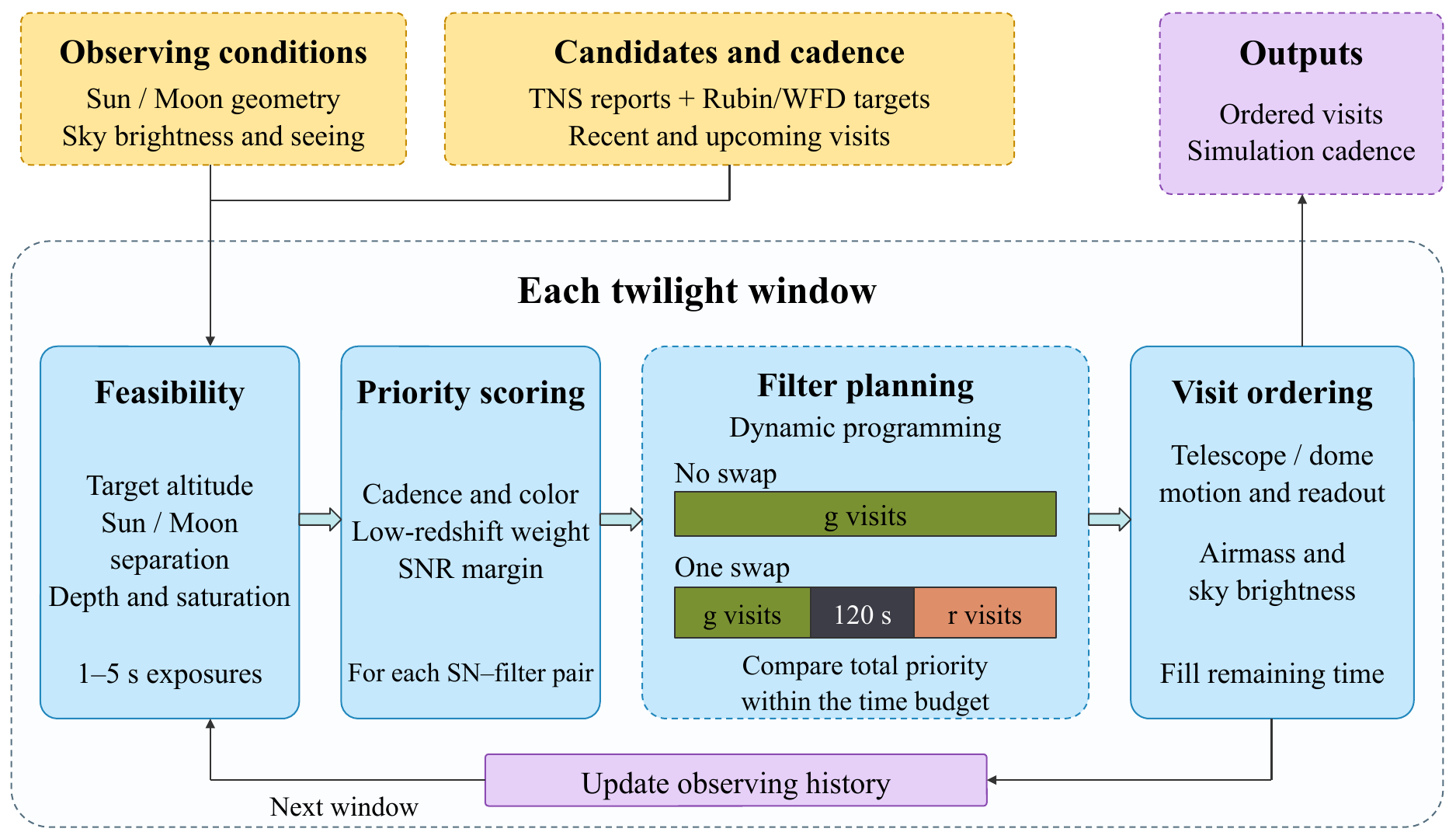}
\caption{\textbf{TW scheduler overview.} For each evening or morning window, the scheduler checks candidate feasibility, assigns priorities to SN--filter pairs, chooses filter intervals, and orders visits. The $g$ and $r$ blocks illustrate two possible filter-order arrangements; this study evaluates all four $g/r/i/z$ filters, and the scheduler can use a user-defined filter set. Filter planning accounts for the observing-time budget and the 120\,s cost of a filter change. Visit ordering accounts for telescope and dome motion, readout, and changing sky conditions. The visit plan supplies the simulation cadence and updates observing history for later windows.}
\label{fig:twilight_overview}
\end{figure}

To coordinate observations, we developed a TW scheduler\footnote{Code repository: \url{https://github.com/trivialTZ/Lsst-Twilight-SNe}.} that constructs visit plans from nightly Sun/Moon ephemerides, a candidate catalog that combines transients reported through TNS with WFD targets (duplicates removed), and (optionally) the expected schedule of upcoming WFD visits.
The scheduler uses \texttt{rubin\_sim} v2.3.0 \citep{rubin_sim_230} for the sky-brightness calculation and \texttt{rubin\_scheduler} v3.15.1 \citep{rubin_scheduler_3151} for the jerk-limited dome-motion model.
For the forecasts, the scheduler also writes an observing-cadence library (SIMLIB) containing the exposure times, filters, and photometric metadata used to generate the simulated light curves. A workflow example is shown in Figure~\ref{fig:twilight_overview}. We simulate a contiguous 365-day interval drawn from survey year~2 of the adopted WFD cadence, after the initial template-acquisition year; difference-imaging templates are therefore assumed to be available across the WFD footprint during the simulated interval.

In real operations, the scheduler generates each night's plan in advance from the current candidate list and observing history. Evening and morning windows are scheduled separately based on solar altitude, $\mathrm{Alt}_\odot$; the main forecast uses $-12^\circ \le \mathrm{Alt}_\odot \le -9^\circ$ together with a topocentric Sun--target separation $\Delta_\odot\ge90^\circ$. Visits default to 5\,s exposures; a saturation check based on the predicted central-pixel charge shortens the exposure (to as little as 1\,s) for the lowest-redshift targets that would otherwise saturate. In the one-year main plan this exposure check affects about 4\% of scheduled visits, while the remaining 96\% use the full 5\,s, so the program operates essentially as a fixed short-exposure survey. The check is validated with DP1 data and simulations (Appendices~\ref{sec:psf_validation} and~\ref{sec:sat_validation}). We choose this darker half of nautical twilight because it preserves useful \emph{griz} depth for 1--5\,s exposures while avoiding brighter twilight, where operations are more constrained; it also gives a repeatable baseline of about 15\,min per window for the forecast.

For each 15-minute window, the time budget includes 120\,s per filter swap, 2\,s of readout per exposure, buffer time to keep exposure starts at least 15\,s apart, and the longer of the telescope-slew and dome-rotation times because the two motions can occur in parallel \citep{RubinDMTN065,RubinTelescopes2026}. The scheduler applies altitude and Sun/Moon separation requirements, with the target airmass entering the photometric predictions. For each proposed exposure, it predicts the signal-to-noise ratio (SNR) and the single-visit $5\sigma$ point-source limiting magnitude $m_5$, and checks the central-pixel charge for saturation. The $90^\circ$ Sun--target separation is checked at the start, midpoint, and end of each exposure. Because twilight sky brightness varies with angular position relative to the Sun, the limiting depth is calculated separately for each target. Appendix~\ref{app:m5_zenith_sky} illustrates this dependence for a 30\,s reference exposure; scheduled-visit depths use each target's coordinates, time, and exposure length. Appendix~\ref{sec:sun_dome_policy} gives the full implementation and its validation.
In the forecast, the scheduler reads the expected WFD visit times by target and filter. The scheduler rejects a TW visit if a same-filter WFD visit is scheduled within the adopted minimum interval before or after it. Because this timing check ignores saturation, it can occasionally reject a TW visit that would have provided an unsaturated measurement near a saturated WFD epoch. For each proposed TW visit, the planner computes airmass and Sun/Moon geometry from the planned time and coordinates. It predicts $m_5$ from the airmass, direction-dependent sky background, assumed per-band seeing and throughput, and exposure length, then uses the estimated target magnitude to predict SNR.

These operational constraints motivate a four-stage pipeline summarized below, with algorithmic details deferred to Appendix~\ref{app:scheduler}.

\begin{samepage}
\paragraph{(1) Feasibility.}
For each candidate SN--filter pair, we require:
\begin{enumerate}
\item target altitude above the configured minimum at some time within the twilight window;
\item an exposure within the adopted solar-altitude window, with topocentric $\Delta_\odot\ge90^\circ$ at its start, midpoint, and end;
\item sufficient Moon separation under the altitude/phase rule in Appendix~\ref{sec:filtering};
\item predicted $\mathrm{SNR}\ge5$ at the proposed exposure length;
\item predicted central-pixel charge below the saturation threshold.
\end{enumerate}
\end{samepage}
If the nominal 5\,s exposure fails the saturation check, the planner shortens it as needed, subject to a 1\,s minimum, and reevaluates feasibility.

\paragraph{(2) Priority scoring.}
For each feasible SN--filter pair, we rank candidates by observing need and per-filter cadence, favor a missing recent color band, and increase priority toward low redshift. In the forecast configuration, a smooth redshift boost acts below $z=0.08$, with an additional factor of 1.8 for SN Ia-like targets below that redshift. These priorities favor the nearest events, including those at $z<0.06$, but do not guarantee a volume-limited sample. The optional SNR-margin factor is disabled in the reported forecast.

\paragraph{(3) Filter planning.}

Observing targets in strict priority order is inefficient because a single carousel swap ($\sim$120\,s) consumes roughly 13\% of a 15\,min window. A dynamic program selects one or a few filter intervals for the window while including the $\sim120$\,s swap penalty in the time budget (optionally imposing a swap cap or requiring a minimum fractional gain before allowing a swap; or enforcing a single filter per window). This preserves band diversity without introducing unnecessary filter changes.

\paragraph{(4) Visit ordering.}
Within each chosen filter segment, we order targets using a timing-cost objective that weighs priority per unit observing time against telescope/dome movement overheads and the evolving, direction-dependent airmass and sky brightness across the twilight window.

\paragraph{Outputs.}
We provide a summary of visit totals for each window and a time-ordered list of individual visits. For simulations, the output also includes a cadence with the seeing, sky brightness, and zeropoints used in the SNR/$m_5$ calculation.

\subsection{Survey Simulation}
\label{sec:snana}

We simulate SN~Ia light curves with the SuperNova ANAlysis software (\texttt{SNANA}\footnote{\url{https://github.com/RickKessler/SNANA}}; \citealt{kessler2009snana}), using configurations adapted from a \texttt{Pippin} workflow\footnote{\url{https://github.com/dessn/Pippin}} \citep{Hinton2020_PIPPIN}. The simulation specifies a source population, an observing cadence, and a photometric-noise model. For the low-$z$ WFD and TW samples, we use the near-infrared extension of SALT3 (Spectral Adaptive Light-curve Template~3), SALT3.P22-NIR \citep{Kenworthy2021_SALT3,Pierel2022SALT3NIR,Taylor2023_SALT3update}. Its wavelength coverage supports the observer-frame Rubin $griz$ bands throughout the simulated redshift range. Light-curve stretch and color $(x_1,c)$ are sampled from the DES-SN5YR low-redshift population distributions, including host-mass dependence \citep{Popovic2023_Dust2Dust,Sanchez2024_DES5YRrelease}.

Each SN is assigned to a redshift-matched host from a Roman--Rubin Diffsky simulated galaxy library \citep{OpenUniverse2025}. The host stellar mass determines the SN and dust population distributions. Galactic extinction follows the \citet{SchlegelFinkbeinerDavis1998} map with the \citet{SchlaflyFinkbeiner2011} recalibration and an $R_V=3.1$ \citet{Fitzpatrick1999} extinction law. At each observation, the simulation converts model flux into CCD photoelectrons and computes photometric uncertainty from source, sky, host-galaxy, and read noise. The host contribution is evaluated from the galaxy photometry and S\'ersic profile at the SN position over the PSF noise-equivalent area. For TW scheduling, predicted source, local-host, and sky photoelectrons in the central pixel are compared with $N_{\rm sat}=8\times10^4\,\mathrm{e}^-$; the exposure is shortened when necessary (Section~\ref{sec:filtering}). Read noise enters the uncertainty but does not add charge to this threshold. The sample-count tables, light-curve diagnostics, and dark-energy forecast use light-curve fits without epoch-level saturation rejection. For the PV forecast, we repeat the low-redshift simulations with identical observing schedules, SN realizations, and random seed, enabling the $8\times10^4\,\mathrm{e}^-$ central-pixel threshold and excluding flagged epochs before applying the same quality cuts. The fixed higher-redshift WFD component is unchanged. Appendix~\ref{sec:sat_validation} presents a separate bright-limit comparison with commissioning data.

We simulate three low-redshift observing cases: baseline WFD observations (\sampleWFD{}), WFD observations with TW epochs appended for the same SNe~Ia (\sampleTWWFD{}), and TW follow-up of additional targets predominantly reported by TNS (\sampleTWTNS{}). The latter is disjoint from the WFD parent sample. The WFD observations come from a contiguous 365-day interval in survey year~2 of the \texttt{baseline\_v4.3\_10yrs} OpSim cadence; this window inherits the rolling cadence pattern present in that year. Each SN in the matched WFD cases retains the same simulated source properties, with the scheduled TW observations added to its WFD light curve in the \sampleTWWFD{} case. The WFD parent sample is drawn from a baseline simulated realization without applying an additional search-efficiency trigger cut. The simulation assumes classified SNe~Ia and measured host redshifts for candidate follow-up. Section~\ref{sec:sample-definitions} defines the quality-selected samples.

To evaluate dark-energy sensitivity, we also include simulated WFD and Deep Drilling Field (DDF) comparison samples at higher redshift. These fixed components use the SALT3 surface released with DES-SN5YR \citep{Sanchez2024_DES5YRrelease}; their redshift distributions and aggregate inverse-variance precision are identical in the baseline and TW-augmented Fisher calculations. The fixed WFD component enters only at $z\ge0.11$. The DDF component is a quality-selected one-year sample spanning approximately $61149\le\mathrm{PKMJD}<61515$; it contains 5058 SNe over $0.0287\le z\le1.4913$ and is retained over its full redshift range in both forecasts. Table~\ref{tab:sample_defs_counts} summarizes these fixed components.

\subsection{Sample Definitions}
\label{sec:sample-definitions}

Table~\ref{tab:sample_defs_counts} defines the low-redshift samples after the common quality cuts in Section~\ref{sec:cosmo-hubble} and lists the fixed WFD and DDF comparison samples. TW follow-up adds 230 selected SNe in the disjoint \sampleTWTNS{} component and changes the selection of the matched WFD parent sample. After light-curve fitting, \sampleTWWFD{} contains 1102 selected SNe at $z<0.11$, compared with 1005 in \sampleWFD{}: 108 enter and 11 leave the selected sample, a net increase of 97. For 1398 matched light-curve fits, the median fitted color uncertainty decreases from 0.02420 in \sampleWFD{} to 0.02144 in \sampleTWWFD{}; added phase and band coverage can also change whether an event passes the fitted peak-date and stretch uncertainty cuts. We define \sampleCombined{} as the union of \sampleTWWFD{} and the disjoint \sampleTWTNS{} component. The main nearby science sample is at $z<0.10$; the dark-energy forecast replaces the WFD component below $z=0.11$, while the PV forecast extends to $z<0.14$ as described in Section~\ref{sec:pv-forecast}.

\begin{table*}[!htbp]

\caption{One-year sample definitions and selected counts\label{tab:sample_defs_counts}}
\centering
\scriptsize
\setlength{\tabcolsep}{3.5pt}
\renewcommand{\arraystretch}{1.18}
\begingroup
\newcommand{\samplecell}[2]{\parbox[t]{#1\textwidth}{\raggedright\strut #2\strut}}
\begin{tabular}{@{}llll@{}}
\hline\hline
\samplecell{0.18}{Sample} & \samplecell{0.33}{Definition} & \samplecell{0.28}{Selected count\tablenotemark{a}} & \samplecell{0.14}{Candidate source} \\
\hline
\samplecell{0.18}{\sampleWFD{}} & \samplecell{0.33}{WFD observations of the parent sample} & \samplecell{0.28}{149 / 751 / 1005} & \samplecell{0.14}{Rubin/WFD} \\
\samplecell{0.18}{\sampleTWWFD{}} & \samplecell{0.33}{Same SNe with WFD and TW observations; refit and reselected} & \samplecell{0.28}{165 / 827 / 1102} & \samplecell{0.14}{Rubin/WFD} \\
\samplecell{0.18}{\sampleTWTNS{}} & \samplecell{0.33}{Disjoint additional TW targets} & \samplecell{0.28}{193 / 228 / 228} & \samplecell{0.14}{Predominantly TNS} \\
\samplecell{0.18}{\sampleCombined{}} & \samplecell{0.33}{Union of \sampleTWWFD{} and \sampleTWTNS{}} & \samplecell{0.28}{358 / 1055 / 1330} & \samplecell{0.14}{Both streams} \\
\hline
\samplecell{0.18}{WFD (higher-$z$; fixed)} & \samplecell{0.33}{Fixed WFD comparison sample at $z\ge0.11$} & \samplecell{0.28}{78,410\tablenotemark{b}} & \samplecell{0.14}{Rubin/WFD} \\
\samplecell{0.18}{DDF (fixed)} & \samplecell{0.33}{Fixed DDF comparison sample over $0.0287\le z\le1.4913$} & \samplecell{0.28}{5,058\tablenotemark{b}} & \samplecell{0.14}{Rubin/DDF} \\
\hline
\end{tabular}
\endgroup
\tablenotetext{a}{For the four low-redshift samples, counts are cumulative at $z<0.06$, $z<0.10$, and $z<0.11$, respectively, after the common light-curve cuts in Section~\ref{sec:cosmo-hubble}. The \sampleTWTNS{} component contains 230 selected SNe over its full redshift range, including two above $z=0.11$. The WFD pair uses the same parent SNe, and its counts differ because each light curve is fitted and selected separately. PV counts appear in Section~\ref{sec:pv-forecast}.}
\tablenotetext{b}{The fixed WFD and DDF components have the same redshift distributions and aggregate inverse-variance precisions in both dark-energy forecasts. The WFD count is the sum of the retained $z\ge0.11$ sample. The DDF component retains its full one-year range, including 15 SNe below $z=0.11$. Its fitted-peak-date window is approximately $61149\le\mathrm{PKMJD}<61515$. These components have their prior selection rather than the new low-$z$ common cuts.}
\end{table*}

\subsection{Analysis and Forecasts}

\label{sec:cosmo}

\subsubsection{Light-Curve Fitting and Distance Estimation}
\label{sec:lcfit}

We fit the simulated light curves with SALT3 using \texttt{SNANA}, obtaining $m_B$, $x_1$, and $c$ and their covariances. To determine a distance modulus for each event, we use the Tripp relation, $\mu=m_B+\alpha x_1-\beta c+\mathcal{M}$ \citep{Tripp1998}, and propagate the full fit covariance with fixed $\alpha=0.14$ and $\beta=3.1$. Here $\mathcal{M}$ sets the reference zero point; a common additive magnitude offset $M$ is marginalized analytically in the Fisher calculation. The per-SN distance uncertainty $\sigma_\mu$, including the additive intrinsic-scatter term obtained from a BEAMS with Bias Corrections analysis (BBC; \citealt{KesslerScolnic2017_BBC}) of simulated DEBASS and DES SNe \citep{acevedo2025dark}, is defined in Section~\ref{sec:cosmo-hubble}.

For the Hubble diagram forecast, the baseline uses the quality-selected \sampleWFD{} events below $z=0.11$; the TW case uses the quality-selected \sampleTWWFD{} events in the same range and adds the disjoint \sampleTWTNS{} events once. The fixed WFD comparison component at $z\ge0.11$ and the full DDF component are identical in both forecasts (Table~\ref{tab:sample_defs_counts}). The Fisher calculation uses each simulated low-$z$ event's assigned redshift and propagated distance uncertainty, together with the retained high-$z$ redshift and precision summaries.

\subsubsection{Forecast Approach and Assumptions}
\label{sec:forecast-assumptions}

We use a Fisher calculation to compare how much information the two simulated samples provide under the same distance-error model and selections. The calculation combines derivatives of the cosmological distance modulus with the summed inverse-variance precision of the selected SNe, as specified in Section~\ref{sec:cosmo-hubble}. We marginalize over the common additive magnitude offset and apply no external cosmological priors. The calculation does not include a full survey-systematics covariance or redshift-dependent selection-bias correction; the resulting parameter errors and figure of merit are statistical forecasts under the adopted scatter model.

The low-$z$ forecast assumes candidate SNe have reliable Type~Ia classifications and spectroscopic host redshifts. It does not simulate additional losses from alert latency, broker filtering, or spectroscopic incompleteness. These requirements apply to both TNS-reported and WFD-alert candidates and underlie the sample counts in Section~\ref{sec:results}. The distance-error prescription in Section~\ref{sec:cosmo-hubble} includes propagated light-curve fit uncertainties, lensing, and the stated scatter terms.

\subsubsection{Hubble Diagram and Cosmic Expansion Constraint Forecast}
\label{sec:cosmo-hubble}

We assess the cosmological impact of TW observations on the SN Hubble diagram with Fisher forecasts for flat $\Lambda$CDM and $w_0$--$w_a$CDM. Each simulated low-$z$ SN contributes its assigned redshift, fitted light-curve parameters, and fit covariance. The high-$z$ components contribute fixed redshift-bin sufficient statistics. We combine the inverse-variance weights of individual SNe within each redshift bin and evaluate the model derivatives at the inverse-variance-weighted mean redshift of that bin.

\paragraph{Windowing and quality requirements.}
We apply the same cuts to \sampleWFD{}, \sampleTWWFD{}, and \sampleTWTNS{} light-curve fits:

\begin{itemize}
\item SALT3 fit probability $>0.05$, the upper-tail probability of the fitted $\chi^2$ for its reported degrees of freedom;
\item $|x_1|\le3$ and $|c|\le0.3$;
\item $\sigma_{t_{\rm peak}}\le1$\,d, $\sigma_{x_1}\le1$, and $\sigma_c\le0.05$;
\item maximum reported SNR $\ge5$ and total light-curve SNR $\ge20$.
\end{itemize}

The SNR thresholds and peak-date uncertainty cut enforce light-curve sampling quality. The one-year WFD interval is the same survey-year-2 interval specified in Section~\ref{sec:snana}. The fixed WFD and DDF comparison components retain their prior selection, which is identical between the two dark-energy forecasts. At the Fisher stage, the TW case substitutes the fitted \sampleTWWFD{} low-$z$ component for \sampleWFD{} and adds the disjoint \sampleTWTNS{} component once (Table~\ref{tab:sample_defs_counts}).

\paragraph{Binning and per-SN uncertainties.}

For the Fisher calculation, we compress the selected SNe into uniform redshift bins of width $\Delta z{=}0.001$. For each SN we propagate the SALT3 fit covariances to a distance-modulus uncertainty \(\sigma_\mu\) and include standard nuisance contributions, adapting the distance-uncertainty decomposition of \citet{brout2022pantheonplus}:

\begin{equation}
\label{eq:sigmamu_sn}
\sigma_\mu^2 = \sigma_{m_B}^2 + (\alpha\,\sigma_{x_1})^2 + (\beta\,\sigma_{c})^2
 + 2\alpha\,\mathrm{C}_{m_B x_1} - 2\beta\,\mathrm{C}_{m_B c}
 - 2\alpha\beta\,\mathrm{C}_{x_1 c} + \sigma_{\rm lens}^2(z) + \sigma_{\rm pv}^2(z) + \sigma_{\rm int}^2,
\end{equation}

where $\sigma_{m_B}$, $\sigma_{x_1}$, and $\sigma_{c}$ are the SALT3 fit uncertainties and $\mathrm{C}_{xy}$ the corresponding fit covariances, with a weak-lensing scatter term in magnitudes, $\sigma_{\rm lens}(z){=}0.055\,z$, and a peculiar-velocity term

\begin{equation}
\sigma_{\rm pv}(z) = \frac{5}{\ln 10}\,\frac{\sigma_{v\!\,\rm pec}}{c_{\rm light}\,z},
\end{equation}

with fiducial $\sigma_{v\!\,\rm pec}{=}300$\,km\,s$^{-1}$ and speed of light $c_{\rm light}$. The additive intrinsic-scatter term $\sigma_{\rm int}(c)$ represents the simulation-derived component $\sigma_{\rm scat}$ of the standardization floor in Equation~(4) of \citet{brout2022pantheonplus}, following the color and dust model of \citet{BroutScolnic2021_Dust}.
We use a BBC run of simulated DEBASS low-redshift and DES SNe \citep{acevedo2025dark} to estimate this term in local fitted-color windows. We take the square root of the median nonnegative additive variance, smooth the resulting profile, and interpolate in color. The adopted curve spans $0$--$0.168$\,mag over $-0.3\le c\le0.3$ and has $\sigma_{\rm int}(0)=0.066$\,mag; its 61 tabulated values are included with the analysis code.
The same $\sigma_{\rm int}(c)$ curve is used for the Hubble diagram and PV forecasts.

For redshift bin $b$ with $N_b$ SNe, we define $\mathcal{W}_{\mu,b}\equiv\sum_{i=1}^{N_b}\sigma_{\mu,i}^{-2}$ and $\sigma_{\mu,\mathrm{bin},b}=\mathcal{W}_{\mu,b}^{-1/2}$. The equivalent per-SN uncertainty shown in the results is given by $\sigma_{\mu,\mathrm{eff},b}\equiv\sqrt{N_b/\mathcal{W}_{\mu,b}}$, distinct from the uncertainty of the weighted mean. We evaluate the cosmological derivatives at the inverse-variance-weighted mean redshift,

\begin{equation}
    z_{\mathrm{eff},b}=\sum_i z_i\sigma_{\mu,i}^{-2}/\mathcal{W}_{\mu,b}.
\end{equation}

 In this Hubble-diagram forecast, the PV contribution is treated as an uncorrelated magnitude scatter term. In the growth-rate forecast (Section~\ref{sec:pv-forecast}), we instead remove that term from the per-SN measurement noise and model the correlated velocity covariance explicitly.

\paragraph{Model and Fisher construction.}
We assume flat geometry and the Chevallier--Polarski--Linder (CPL) equation of state, $w(z)=w_0+w_a z/(1+z)$ \citep{ChevallierPolarski2001,Linder2003}. The corresponding expansion history is

\begin{equation}
E(z) = \Big[\Omega_m (1{+}z)^3 + (1{-}\Omega_m) (1{+}z)^{3(1+w_0+w_a)}\,\exp\!\big( {-}3w_a z/(1{+}z) \big)\Big]^{1/2},
\end{equation}

with comoving and luminosity distances given by \(D_C(z) = (c/H_0)\int_0^z dz'/E(z')\) and \(D_L(z)=(1{+}z)D_C(z)\), and

\begin{equation}
\mu(z;\,\boldsymbol{p}) = 5\log_{10}\!\big[D_L(z)/\mathrm{Mpc}\big] + 25 + M.
\end{equation}

In these equations $\Omega_m$ is the present-day matter-density fraction, $w_0$ and $w_a$ define the CPL dark-energy equation of state, $H_0$ is the Hubble constant at $z=0$, and $M$ is a common additive magnitude offset that absorbs the unknown absolute distance normalization.

We build a Jacobian $J$ of partial derivatives of $\mu$ with respect to the parameter vector $(\Omega_m,M)$ for $\Lambda$CDM or $(\Omega_m,w_0,w_a,M)$ for $w_0$--$w_a$CDM, evaluated at $z_\mathrm{eff}$ in each bin. With weights $W_{bb}=\mathcal{W}_{\mu,b}$ for redshift bin $b$ on the diagonal of $W$, the full Fisher matrix is $F_{\rm full}{=}J^\mathsf{T} W J$.

We report the SN-only cosmological constraints by analytically marginalizing the additive magnitude offset $M$ via the Schur complement,

\begin{equation}
F_{\rm marg} = F_{\theta\theta} - F_{\theta M}\,F_{MM}^{-1}\,F_{M\theta},
\end{equation}

and inverting $F_{\rm marg}$ for covariances and 1$\sigma$ errors. No external priors are imposed in the baseline forecasts.

\paragraph{Reporting.}
For $\Lambda$CDM and $w_0$--$w_a$CDM we report marginalized parameter errors. For the evolving-dark-energy model we define the figure of merit as the inverse area of the marginalized $w_0$--$w_a$ covariance ellipse up to a common confidence-level factor:

\begin{equation}
    \mathrm{FoM}_{w_0w_a}=[\det C(w_0,w_a)]^{-1/2}.
\end{equation}
These constraints and the figure of merit are reported in Section~\ref{sec:cosmology}.

\subsubsection{Peculiar Velocities and \texorpdfstring{$f\sigma_8$}{fsigma8} Forecast}
\label{sec:pv-forecast}

We construct a peculiar velocity (PV) forecast from simulated SN~Ia samples, using
the saturation-enabled light-curve-fit products described in Section~\ref{sec:snana} and the field-level Fisher forecast implemented in
\texttt{FLIP}\footnote{\url{https://github.com/peculiar-velocity/flip}.} \citep{Ravoux2025_FLIP}.
The baseline PV catalog combines the low-$z$ WFD realization with a high-$z$ WFD component; the TW-augmented catalog is processed identically.

\paragraph{One-year selection and quality control.}

For the one-year PV comparison, using the saturation-enabled low-redshift fits, the baseline combines \sampleWFD{} at $z<0.10$ with a matching fixed WFD simulation component at $0.10\le z<0.14$. The TW-augmented comparison substitutes \sampleTWWFD{} below $z=0.10$ and adds \sampleTWTNS{}, retaining the same fixed WFD component. We evaluate the ranges $0.02\le z<0.06$, $0.06\le z<0.10$, and $0.02\le z<0.14$ with this construction.

We apply the fit-quality and summary-SNR cuts in Section~\ref{sec:cosmo-hubble}, following \citet{fs8_forecast_2507_00157}. In the three ranges above, \sampleWFD{} has 109, 600, and 1834 SNe, respectively; \sampleCombined{} has 296, 695, and 2117. The last range includes the same fixed $0.10\le z<0.14$ WFD component in both cases.

\paragraph{Data vector and distance-redshift mapping.}

For each SN, we compute the distance uncertainty from the full SALT3 fit covariance with $\alpha=0.14$ and $\beta=3.1$, adding the same $\sigma_{\rm int}(c)$ used in Equation~\ref{eq:sigmamu_sn}. We retain the lensing term and omit the uncorrelated peculiar-velocity term from this distance-error input. Sensitivity to $f\sigma_8$ in this Fisher forecast comes from the modeled covariance of correlated peculiar velocities at the simulated sky positions and redshifts.

Each SN enters the forecast with its sky position and observed redshift; we evaluate
the comoving distance and expansion rate with the \texttt{Planck18} cosmology \citep{Planck2018_A6}. We pass \texttt{FLIP} the consistent $h$-scaled pair $r_{\rm com}$ in Mpc$/h$ and $H(z)/h=100\,E(z)$ in km\,s$^{-1}$\,Mpc$^{-1}$.

\paragraph{Covariance, power spectrum, and Fisher forecast.}
We construct the Hubble-diagram residual covariance with the \texttt{carreres23} model in \texttt{FLIP}, combining diagonal measurement terms (including the per-SN $\sigma_{\rm int}(c)$) with large-scale PV. The linear velocity-divergence power spectrum is computed with a CLASS backend \citep{lesgourgues2011_class} (normalized to the growth rate) and damped by a small-scale kernel $D_u(k,\sigma_u)$, where $k$ is wavenumber and $\sigma_u$ is the nonlinear velocity-damping scale; we adopt $\sigma_u=21\,h^{-1}\,\mathrm{Mpc}$ in the \texttt{FLIP} convention.

We forecast the normalized amplitude $A_{f\sigma_8}\equiv(f\sigma_8)/(f\sigma_8)_{\rm fid}$ and form a full-rank Fisher matrix for $\{A_{f\sigma_8},\,\sigma_v,\,s_M\}$, evaluated at $\{1,\,300\,\mathrm{km\,s^{-1}},\,0\,\mathrm{mag}^2\}$. Here $\sigma_v$ is a small-scale velocity-dispersion nuisance term and $s_M\equiv\sigma_{M,\mathrm{add}}^2$ is an added magnitude variance beyond the color-dependent intrinsic scatter already included in the per-SN distance error. We marginalize over $\sigma_v$ and $s_M$ without external priors. The calculation uses $10^{-5}\le k\le1\,h\,\mathrm{Mpc}^{-1}$ and 1500 $k$ samples.

Two selection-related effects are relevant to these forecasts. First, incomplete sky coverage enters explicitly through the window function, since the velocity covariance is constructed from the actual $(\alpha,\delta,z)$ of the simulated sample. Second, flux-limited discovery and heterogeneous spectroscopic follow-up can induce redshift-dependent distance biases. Modeling these selection-induced mean-distance biases requires an explicit multi-survey selection function and is left outside the present forecast. As a separate robustness test for real data, low-order angular nuisance terms (e.g., dipole/quadrupole modes) should be carried in PV fits to absorb residual large-angle calibration or selection patterns.

\paragraph{Reporting.}

For designated low-$z$ ranges, we quote $(N, z_{\rm eff}, \sigma_{f\sigma_8})$ for
\sampleWFD{} and \sampleCombined{} samples, and compare the fractional precision $\sigma_{f\sigma_8}/(f\sigma_8)_{\rm fid}$ as a function of redshift.

\section{\texorpdfstring{Results of the One-Year Survey Simulation and Forecasts}{Results of the One-Year Survey Simulation and Forecasts}}
\label{sec:results}

Below we present the trigger-stream properties, scheduled twilight cadence, low-$z$ Hubble-diagram improvements, and resulting dark-energy and PV forecasts.

\subsection{Low-\texorpdfstring{$z$}{z} SN Sample Distribution and Redshift Coverage}
\paragraph{TNS as an empirical trigger-stream substitute.}
We use a 2022--2023 Transient Name Server (TNS) SN~Ia compilation as an empirical proxy for the external portion of the triggers that a Rubin TW program would follow. TNS is not a homogeneous survey; it aggregates discovery and classification reports from surveys and brokers with different footprints, limiting magnitudes, and follow-up resources \citep{TNS_home,TNS_statsmaps}. In the low-redshift regime relevant here, major contributors include wide-area programs such as ZTF \citep{Bellm2019_ZTF,Fremling2020_ZTFBTS_I} and ATLAS \citep{Tonry2018_ATLAS,Smith2020_ATLAS_TSS}. We use the observed TNS sky and redshift distributions after Rubin-accessibility cuts as a practical external-trigger baseline, not as a complete parent population. During Rubin operations, this stream would be combined with Rubin alerts for nearby events predicted to saturate near peak or to lack adequate peak-phase coverage under the scheduled WFD cadence. The forecasts are conditional on a comparable combined external-and-Rubin trigger stream.

\begin{figure*}
\centering
\includegraphics[width=0.95 \linewidth]{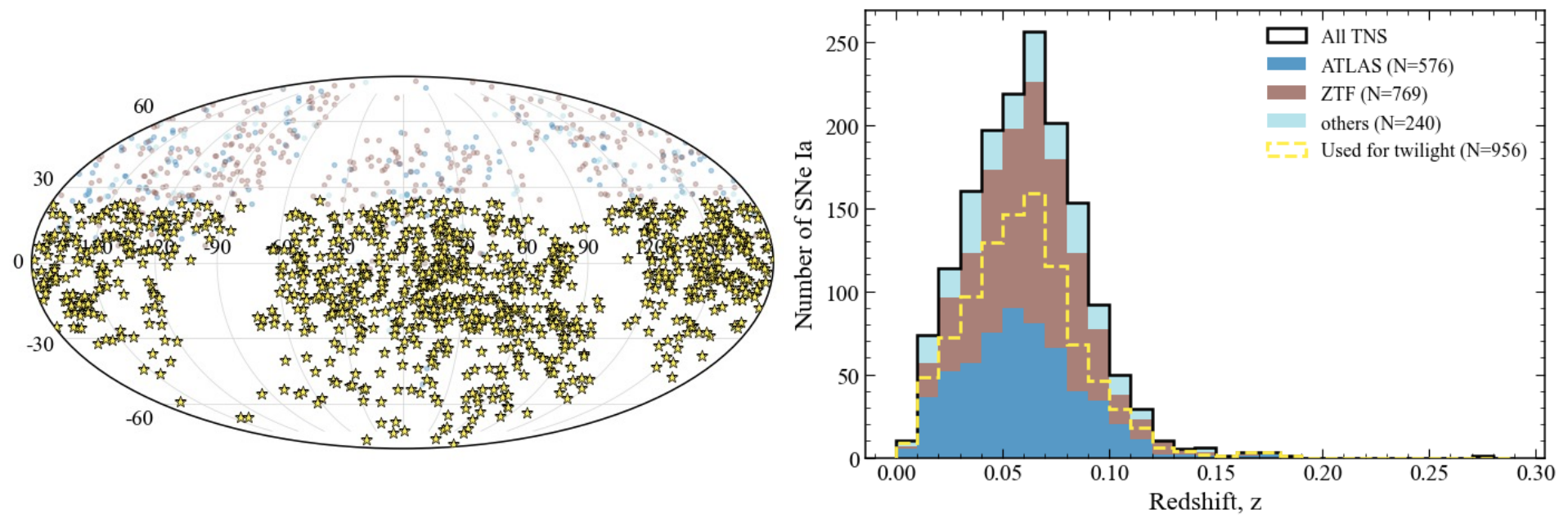}
\caption{\textbf{Empirical low-redshift SN~Ia trigger stream used as scheduler input.} \emph{(a)} Equatorial distribution of SNe~Ia reported to TNS during calendar year 2023, grouped by the discovery stream used in this analysis. Yellow stars mark the subset that passes the scheduler-input cuts. \emph{(b)} Redshift distribution of the TNS comparison sample from November 2022 through December 2023. The dashed yellow histogram shows the twilight trigger subset, which favors Rubin-accessible declinations and follows the low-$z$ shape of the TNS distribution. \label{fig:tns}}
\end{figure*}

Figure~\ref{fig:tns} shows the sky and redshift distributions of the empirical trigger stream before light-curve simulation and cosmology quality cuts. The Rubin-accessible subset is concentrated at declinations $\delta\lesssim +30^\circ$, as expected from the Cerro Pach\'on site latitude, and follows the low-$z$ part of the parent TNS redshift distribution, peaking near $z\sim0.06$ and declining beyond $z\gtrsim0.1$.
Table~\ref{tab:sample_defs_counts} summarizes how these input streams map to the quality-selected forecast samples.

\subsection{TW Survey Coverage and Observing Efficiency}
\begin{figure*}
\centering
\includegraphics[width=\textwidth]{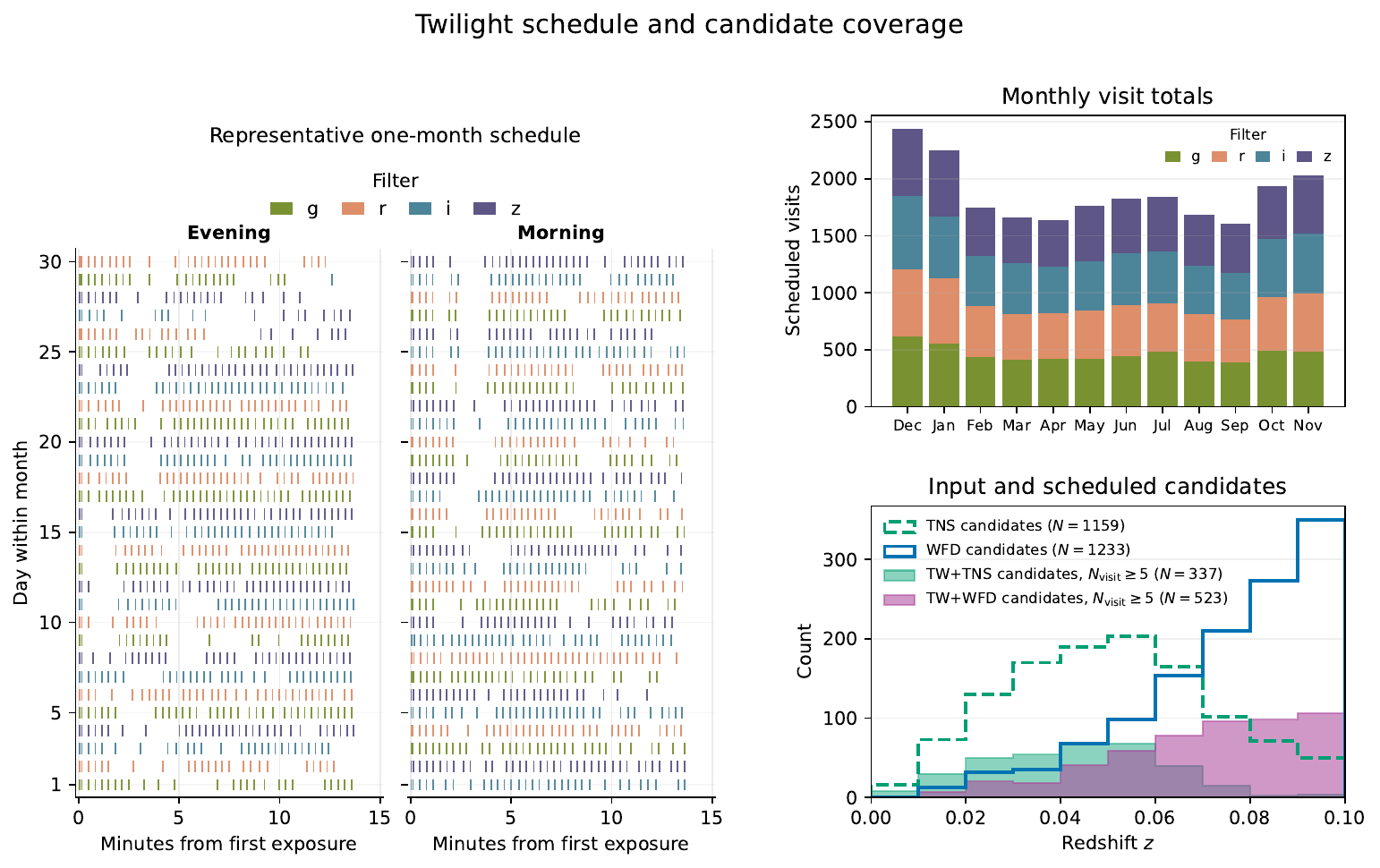}
\caption{\textbf{Twilight schedule and candidate coverage.} The schedule follows the observing policy in Section~\ref{sec:scheduler-overview}. \emph{Left:} Exposure timeline for one representative month; each row is a day, with evening and morning windows shown separately and colors denoting $g/r/i/z$. Horizontal position is minutes from the first exposure in that window. \emph{Upper right:} Scheduled visits by month and filter. \emph{Lower right:} Redshift distributions of TNS and WFD input candidates (outlines) and their respective scheduled subsets with at least five TW visits (filled histograms). These are scheduler-stage counts, before light-curve fitting and cosmology quality cuts.}
\label{fig:twilight_overview_combo}
\end{figure*}

For the policy in Section~\ref{sec:scheduler-overview}, the one-year schedule places 22,416 visits on 1,310 targets. Figure~\ref{fig:twilight_overview_combo} shows when these visits occur, their filter mix, and the redshift distributions of candidates with at least five planned visits. After light-curve fitting and quality cuts, the \sampleTWTNS{} sample contributes 230 SNe, of which 227 are at $z<0.08$; adding TW epochs to WFD light curves gives 97 more selected low-$z$ SNe than WFD alone.

Most planned visits use 5\,s exposures, with about 4\% shortened by the saturation cap. The forecast depends on a twilight sky model fitted only through $\mathrm{Alt}_\odot=-11^\circ$: 14,949 of the 22,416 main-policy visits (66.7\%) occur above that limit and use extrapolated sky brightness. A control schedule confined to $-12^\circ\le\mathrm{Alt}_\odot\le-11^\circ$ contains 6,221 visits and produces 105 quality-selected \sampleTWTNS{} SNe. Appendix~\ref{app:scheduler} details the geometric and slew checks, and Appendix~\ref{sec:sat_validation} presents separate saturation-validation tests. As described in Section~\ref{sec:snana}, epoch-level saturation rejection is disabled in the fits used for the sample counts and dark-energy forecast and enabled in the PV forecast.

\subsection{Light-Curve Quality in TW vs.\ WFD}

\begin{figure*}
\centering
\includegraphics[width=0.98\textwidth]{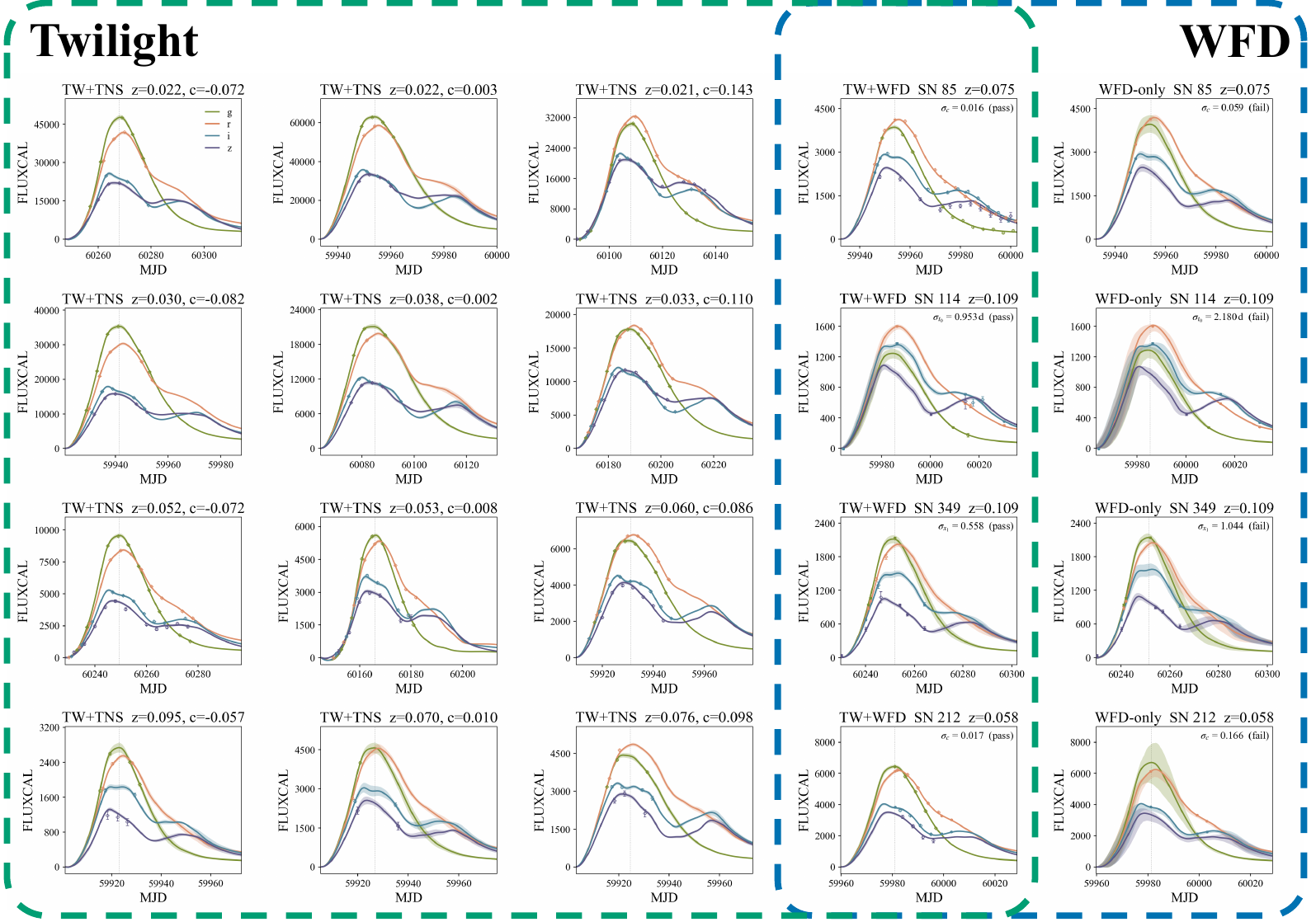}
\caption{\textbf{Representative fitted SN~Ia light curves.} Points show simulated calibrated fluxes at zero point 27.5 with $1\sigma$ measurement errors. Solid curves show SALT3 (P22-NIR) fits without epoch-level saturation rejection; shading encloses the 16th--84th percentiles from draws of the full fitted-parameter covariance. Colors denote Rubin $g/r/i/z$. The left three columns show \sampleTWTNS{} SNe across redshift and fitted color. The right two columns compare the same SN with and without TW observations, using matched axes. Annotations give the fitted uncertainty and quality-cut outcome; the relevant thresholds are $\sigma_c\le0.05$, $\sigma_{t_0}\le1$\,day, and $\sigma_{x_1}\le1$. Open diamonds denote TW observations, filled circles denote WFD observations, and dotted lines mark fitted peak. \label{fig:lc_grid}}
\end{figure*}

Figure~\ref{fig:lc_grid} shows representative fitted SN~Ia light curves in Rubin $g/r/i/z$. The first three columns show externally triggered \sampleTWTNS{} objects. The final two columns compare the same WFD-linked SN before and after TW epochs are appended, highlighting the added phase and band coverage.

The paired examples in the final two columns fail the \sampleWFD{} quality cuts and pass after TW observations are added. The first three rows illustrate improvements in color, peak-date, and stretch precision, respectively; the fourth row illustrates improvement across these quantities. For all 1398 WFD-linked objects with fits in both simulations at $z<0.11$, the median fitted color uncertainty decreases from $0.02420$ to $0.02144$, and the median peak-date uncertainty decreases from $0.1943$ to $0.1700$\,d. These medians describe paired fit outputs before the final quality cuts.

\subsection{\texorpdfstring{Low-Redshift Sample Growth and Distance Precision}{Low-Redshift Sample Growth and Distance Precision}}

\begin{figure*}

\gridline{\includegraphics[width=0.49\textwidth]{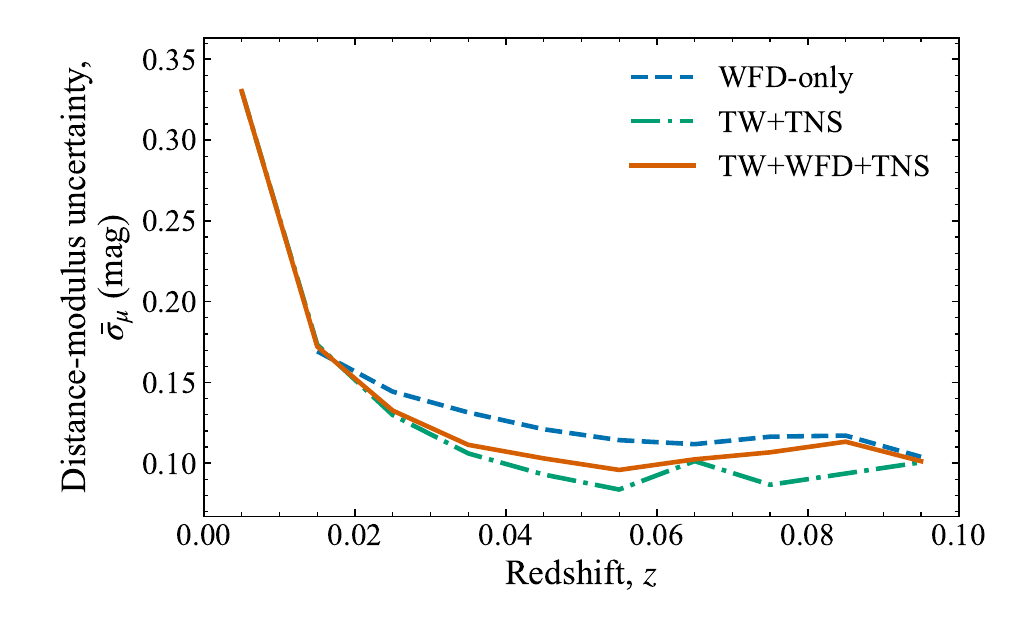} \hfill
          \includegraphics[width=0.49\textwidth]{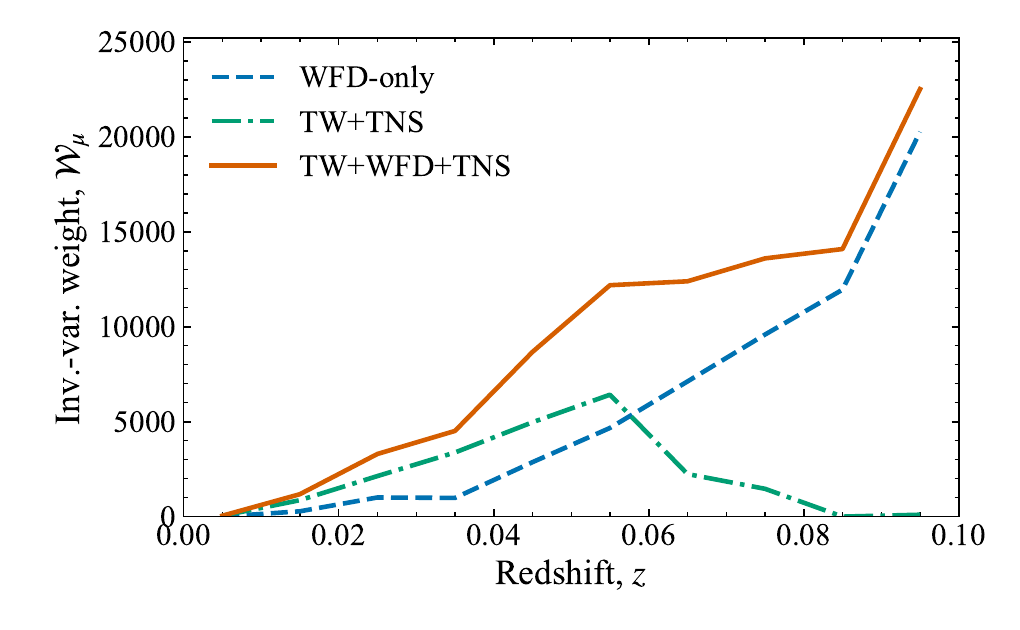}}
\caption{\textbf{Binned Hubble-diagram precision with and without TW.} Both panels show the $z\le0.10$ subset of the one-year catalogs in Tables~\ref{tab:snstats} and~\ref{tab:lowz_wfd_twilight_enhanced}, where the twilight contribution is largest. These catalogs were fitted without epoch-level saturation rejection. \emph{(a)} Equivalent per-SN distance-modulus uncertainty, $\sigma_{\mu,\mathrm{eff}}$ (mag); lower values mean tighter average distances in the bin. \emph{(b)} Inverse-variance weight, $\mathcal{W}_{\mu}=\sum_i1/\sigma_{\mu,i}^2$; larger values mean more Fisher weight. The combined \sampleCombined{} curve is recomputed from the union of the disjoint \sampleTWWFD{} and \sampleTWTNS{} components, so it is not the arithmetic sum of the displayed WFD and \sampleTWTNS{} curves. \label{fig:sigmamu}}
\end{figure*}

Figure~\ref{fig:sigmamu} shows the equivalent per-SN distance uncertainty $\sigma_{\mu,\mathrm{eff}}\equiv\sqrt{N/\mathcal W_\mu}$ and total inverse-variance weight $\mathcal W_\mu=\sum_i1/\sigma_{\mu,i}^2$ for each redshift bin. Table~\ref{tab:snstats} gives corresponding binned counts and distance uncertainties.

The largest count gain occurs below $z=0.05$: the selected sample rises from 88 \sampleWFD{} SNe to 246 in \sampleCombined{} (Table~\ref{tab:snstats}). At $0.01\le z<0.02$, the total weight is 4.23 times the \sampleWFD{} value (Figure~\ref{fig:sigmamu}). The combined sample also has lower equivalent per-SN distance uncertainty than \sampleWFD{} through most populated bins below $z=0.10$. In the same-object comparison, the 742 WFD-linked SNe selected in both fits at $z<0.10$ gain 9.5\% in total inverse-variance weight after TW epochs are added. The combined weight gain therefore comes from both additional selected SNe and tighter distance uncertainties on shared WFD events. The three \sampleTWTNS{} SNe above $z=0.08$ are listed in Table~\ref{tab:snstats}.

\begin{table*}
\caption{One-Year SN~Ia Sample by Redshift: WFD vs.\ \sampleTWTNS{} vs.\ \sampleCombined{}\label{tab:snstats}}
\centering
\begin{tabular}{lcccccc}
\hline\hline
Redshift range & \multicolumn{2}{c}{\sampleWFD{}} & \multicolumn{2}{c}{\sampleTWTNS{}} & \multicolumn{2}{c}{\sampleCombined{}} \\
 & $N$ & $\sigma_{\mu,\mathrm{eff}}$ (mag) & $N$ & $\sigma_{\mu,\mathrm{eff}}$ (mag) & $N$ & $\sigma_{\mu,\mathrm{eff}}$ (mag) \\
\hline
$[0.00,\,0.01)$ & 0   & \nodata & 5   & 0.330 & 5   & 0.330 \\
$[0.01,\,0.02)$ & 8   & 0.169 & 26  & 0.173 & 35  & 0.172 \\
$[0.02,\,0.03)$ & 21  & 0.144 & 36  & 0.130 & 58  & 0.133 \\
$[0.03,\,0.04)$ & 17  & 0.131 & 38  & 0.106 & 56  & 0.111 \\
$[0.04,\,0.05)$ & 42  & 0.121 & 43  & 0.093 & 92  & 0.103 \\
$[0.05,\,0.06)$ & 61  & 0.114 & 45  & 0.084 & 112 & 0.096 \\
$[0.06,\,0.07)$ & 89  & 0.112 & 23  & 0.101 & 130 & 0.102 \\
$[0.07,\,0.08)$ & 130 & 0.116 & 11  & 0.087 & 155 & 0.107 \\
$[0.08,\,0.09)$ & 164 & 0.117 & 0   & \nodata & 181 & 0.113 \\
$[0.09,\,0.10)$ & 219 & 0.104 & 1   & 0.101 & 231 & 0.101 \\
$[0.10,\,0.11)$ & 254 & 0.111 & 0   & \nodata & 275 & 0.109 \\
$[0.11,\,0.12)$ & 350 & 0.145 & 0   & \nodata & 350 & 0.145 \\
$[0.12,\,0.13)$ & 380 & 0.131 & 0   & \nodata & 380 & 0.131 \\
$[0.13,\,0.14)$ & 630 & 0.147 & 1   & 0.106 & 631 & 0.147 \\
$[0.14,\,0.15)$ & 560 & 0.152 & 1   & 0.087 & 561 & 0.152 \\
\hline
\end{tabular}
\tablecomments{Contiguous $\Delta z{=}0.01$ bins from $z{=}0.00$ to $0.15$. Columns list counts ($N$) and equivalent per-SN distance uncertainty in mag ($\sigma_{\mu,\mathrm{eff}} \equiv \sqrt{N/\mathcal{W}_\mu}$), where $\mathcal{W}_\mu \equiv \sum_i 1/\sigma_{\mu,i}^2$, for the baseline WFD sample, the \sampleTWTNS{} sample, and the combined \sampleCombined{} sample (\sampleTWTNS{} + \sampleTWWFD{}) in one year. Larger $\mathcal{W}_\mu$ gives more Fisher weight in that redshift bin. Of the 230 \sampleTWTNS{} SNe, 227 lie at $z<0.08$; the three higher-redshift objects are shown explicitly.}
\end{table*}

Table~\ref{tab:lowz_wfd_twilight_enhanced} gives the WFD-linked comparison separately: \sampleTWWFD{} appends TW epochs to the same simulated objects and refits their light curves. The quality-selected count rises from 1005 to 1102, a net gain of 97 events over $z<0.11$. The disjoint \sampleTWTNS{} population appears in Table~\ref{tab:snstats}.

\begin{table*}
\caption{Low-$z$ WFD sample: \sampleWFD{} vs.\ \sampleTWWFD{} (1 yr)\label{tab:lowz_wfd_twilight_enhanced}}
\centering
\small
\setlength{\tabcolsep}{5.5pt}
\renewcommand{\arraystretch}{1.12}

\begin{tabular}{l
                r r r@{.}l
                r r r@{.}l}
\hline\hline
& \multicolumn{4}{c}{\sampleWFD{}} & \multicolumn{4}{c}{\sampleTWWFD{}} \\
\cline{2-5}\cline{6-9}
Redshift range
& $N_{\rm WFD}$ & $\sigma_{\mu,\mathrm{eff}}$ & \multicolumn{2}{c}{$\mathcal{W}_{\mu}$}
& $N_{\rm WFD}$ & $\sigma_{\mu,\mathrm{eff}}$ & \multicolumn{2}{c}{$\mathcal{W}_{\mu}$} \\
\hline
$[0.00,\,0.01)$ & 0   & \nodata & 0    & 00 & 0   & \nodata & 0    & 00 \\
$[0.01,\,0.02)$ & 8   & 0.169 & 279  & 15 & 9   & 0.169 & 314  & 78 \\
$[0.02,\,0.03)$ & 21  & 0.144 & 1008 & 37 & 22  & 0.137 & 1164 & 25 \\
$[0.03,\,0.04)$ & 17  & 0.131 & 983  & 97 & 18  & 0.126 & 1136 & 17 \\
$[0.04,\,0.05)$ & 42  & 0.121 & 2866 & 92 & 49  & 0.115 & 3703 & 77 \\
$[0.05,\,0.06)$ & 61  & 0.114 & 4670 & 37 & 67  & 0.108 & 5769 & 32 \\
$[0.06,\,0.07)$ & 89  & 0.112 & 7120 & 58 & 107 & 0.103 & 10156 & 10 \\
$[0.07,\,0.08)$ & 130 & 0.116 & 9591 & 03 & 144 & 0.109 & 12139 & 98 \\
$[0.08,\,0.09)$ & 164 & 0.117 & 11956 & 59 & 181 & 0.113 & 14097 & 57 \\
$[0.09,\,0.10)$ & 219 & 0.104 & 20271 & 51 & 230 & 0.101 & 22425 & 83 \\
$[0.10,\,0.11)$ & 254 & 0.111 & 20601 & 10 & 275 & 0.109 & 23071 & 30 \\
\hline
\end{tabular}

\vspace{3pt}
\tablecomments{The ``\sampleWFD{}'' low-$z$ sample corresponds to the baseline WFD simulation.
``\sampleTWWFD{}'' corresponds to the same low-$z$ WFD simulation with additional TW epochs appended and refit.
The appended epochs increase the QC-selected low-$z$ sample (1005$\rightarrow$1102) and increase the inverse-variance weight $\mathcal{W}_{\mu} \equiv \sum_i 1/\sigma_{\mu,i}^2$ in most bins by improving cadence and band coverage for WFD targets.}
\end{table*}

\subsection{Cosmological Parameter Constraints: Dark Energy and Growth Rate}
\label{sec:cosmology}
We next propagate the low-$z$ changes into the two forecast calculations defined in Section~\ref{sec:cosmo}: an SN-only Hubble-diagram Fisher forecast for dark energy and a field-level PV Fisher forecast for $f\sigma_8$.

For the one-year WFD+DDF Hubble-diagram forecast, adding the TW low-$z$ sample increases the $(w_0,w_a)$ FoM from 49.3 to 58.3, an $18.2\%$ gain. The marginalized uncertainties decrease by $7.6\%$ for $\Omega_{\rm m}$ and $10.0\%$ for $w_a$; $\sigma(w_0)$ changes by less than 0.1\%. The high-redshift WFD and DDF components are held fixed, and no external cosmological probes are added. In the calibrated $-12^\circ$ to $-11^\circ$ control, the FoM instead increases from 49.3 to 52.9 ($7.2\%$), quantifying the sensitivity to the extrapolated brighter-twilight portion of the main policy.

We estimate $\sigma(f\sigma_8)/(f\sigma_8)_{\rm fid}$ with the field-level Fisher forecast implemented in \texttt{FLIP} \citep{Ravoux2025_FLIP}, using the redshift range $z\ge0.02$ to avoid the regime where local-flow modeling and redshift-space selection become especially sensitive to nonlinear effects \citep{carreres2023_ztf_growth,fs8_forecast_2507_00157}. The two narrow bins, $0.02\le z<0.06$ and $0.06\le z<0.10$, and the cumulative $0.02\le z<0.14$ range are direct one-year WFD versus \sampleCombined{} comparisons built from the matched saturation-enabled low-redshift realizations, the same fixed higher-redshift component, and common quality cuts.

In the lowest bin, $0.02\le z<0.06$, the one-year WFD forecast gives $\sigma(f\sigma_8)/(f\sigma_8)_{\rm fid}=0.783$ with 109 SNe. \sampleCombined{} gives 0.305 with 296 SNe, a 61.0\% reduction in fractional uncertainty. In $0.06\le z<0.10$, the uncertainty changes from 0.523 to 0.430, a 17.8\% reduction. Across $0.02\le z<0.14$, it changes from 0.302 to 0.215, a 28.7\% reduction. All quoted uncertainties are marginalized. In the calibrated control, the cumulative $0.02\le z<0.14$ uncertainty is 0.249, a 17.5\% reduction relative to WFD.

\begin{figure*}[t]
\centering
\begin{minipage}[t]{0.47\textwidth}
\centering
\includegraphics[width=\linewidth]{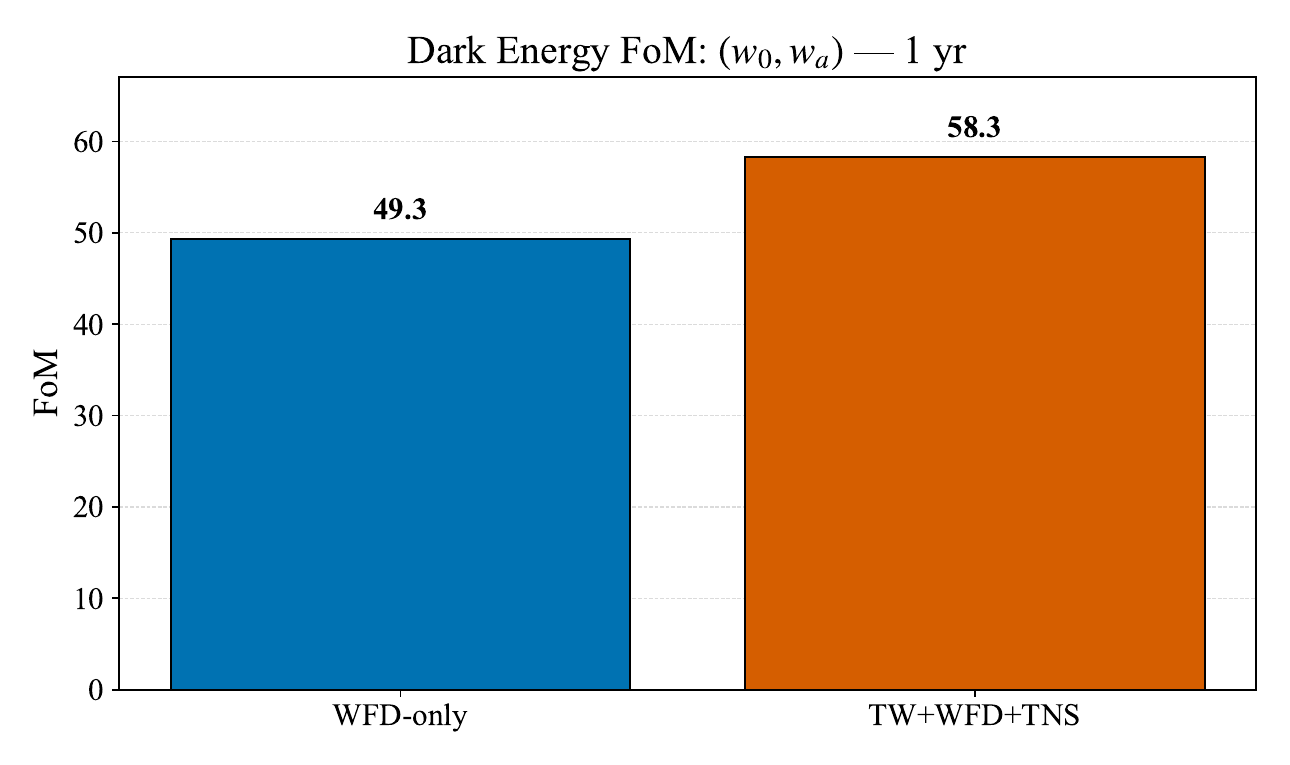}
\end{minipage}
\hfill
\begin{minipage}[t]{0.47\textwidth}
\centering
\includegraphics[width=\linewidth]{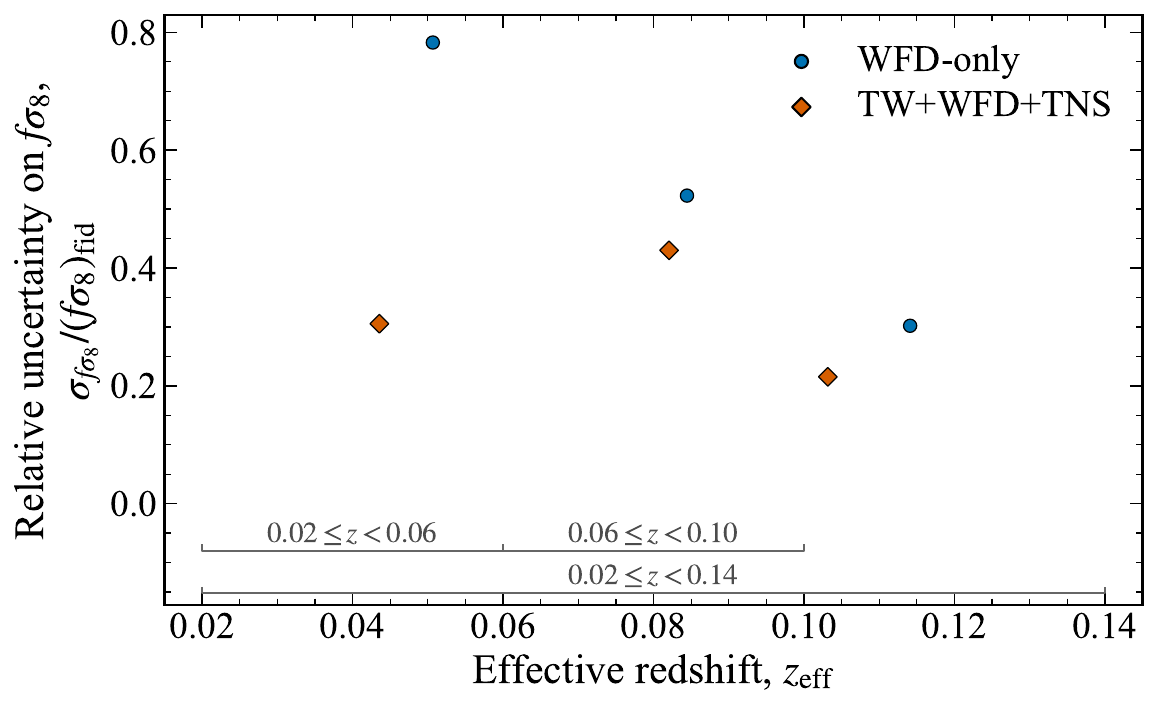}
\end{minipage}
\caption{\textbf{One-year, statistical-only cosmology forecasts.} \emph{Left:} SN-only $(w_0,w_a)$ figure of merit (FoM), with the same fixed high-redshift WFD and Deep Drilling Field samples in both cases. The low-redshift \sampleCombined{} sample increases the FoM from 49.3 to 58.3 using fits without epoch-level saturation rejection. \emph{Right:} Marginalized fractional uncertainty on $f\sigma_8$ from SN peculiar velocities for \sampleWFD{} and \sampleCombined{}, using fits with epoch-level saturation rejection enabled. The lowest-bin uncertainty decreases from 0.783 to 0.305. Gray brackets show the redshift ranges. The two lower-redshift points for each sample represent $0.02\le z<0.06$ and $0.06\le z<0.10$; the point at the highest effective redshift represents the cumulative $0.02\le z<0.14$ sample. Both forecasts include their stated statistical-scatter prescriptions but no calibration-systematics covariance or external cosmological prior. \label{fig:cosmo_forecasts}}
\end{figure*}

\section{Discussion}
\label{sec:discussion}
A one-year TW program substantially expands the nearby SN~Ia sample on Rubin's photometric system, increasing the statistical, SN-only dark-energy FoM by 18.2\% and reducing the $0.02\le z<0.06$ PV fractional uncertainty on $f\sigma_8$ by 61.0\% under our baseline assumptions (with a 7.2\% FoM gain when restricted to the calibrated $-12^\circ\le\mathrm{Alt}_\odot\le-11^\circ$ sky-model domain).

\subsection{Twilight Observing Budget and Operational Constraints}
\label{sec:tw_budget}

Our main window is the darker half of nautical twilight, $-12^\circ \leq \mathrm{Alt}_\odot \leq -9^\circ$, giving about 15\,min per window, or about 30\,min per usable night when both evening and morning windows are available. We apply a $90^\circ$ solar exclusion throughout this window. Of the scheduled visits, 66.7\% occur above $\mathrm{Alt}_\odot=-11^\circ$, outside the fitted range of the twilight sky model. The $-12^\circ$ to $-11^\circ$ control provides a benchmark within the sky model's fitted range, and the predicted gain motivates on-sky validation of the brighter interval. The forecast combines evening and morning observations. The two windows may contribute different numbers of selected SNe because target visibility and sky brightness differ. In the one-year schedule, the airmass at exposure start has a median $X=1.31$, a central 68\% range of $1.08$--$1.60$, a 95th percentile of 1.72, and a maximum of 1.87. The forecast also assumes that a combined external-and-Rubin low-$z$ trigger stream remains available. The effect of incomplete reporting, classification, or alert selection requires an explicit trigger-selection model. Any remaining buffer time within a window absorbs visit-to-visit execution jitter, short broker delays, and slew or settle variations. Weather and engineering interruptions can remove substantial parts of a window or an entire window; their annual effect requires an observing-duty-cycle model.

\subsection{What Very Low Redshift Adds to the Forecasts}
\label{sec:cosmo_power}

In $0.02\le z<0.06$, one year of WFD provides 109 SNe, while the combined sample provides 296; the marginalized $\sigma(f\sigma_8)/(f\sigma_8)_{\rm fid}$ decreases from 0.783 to 0.305. The dark-energy calculation uses the full WFD+DDF Hubble diagram across $0.01\lesssim z\lesssim1.5$. Adding the TW observations raises its two-parameter $(w_0,w_a)$ figure of merit from 49.3 to 58.3 by increasing the statistical weight at the low-redshift end of the Hubble diagram.

\begin{figure*}
\centering
\includegraphics[width=0.98\textwidth]{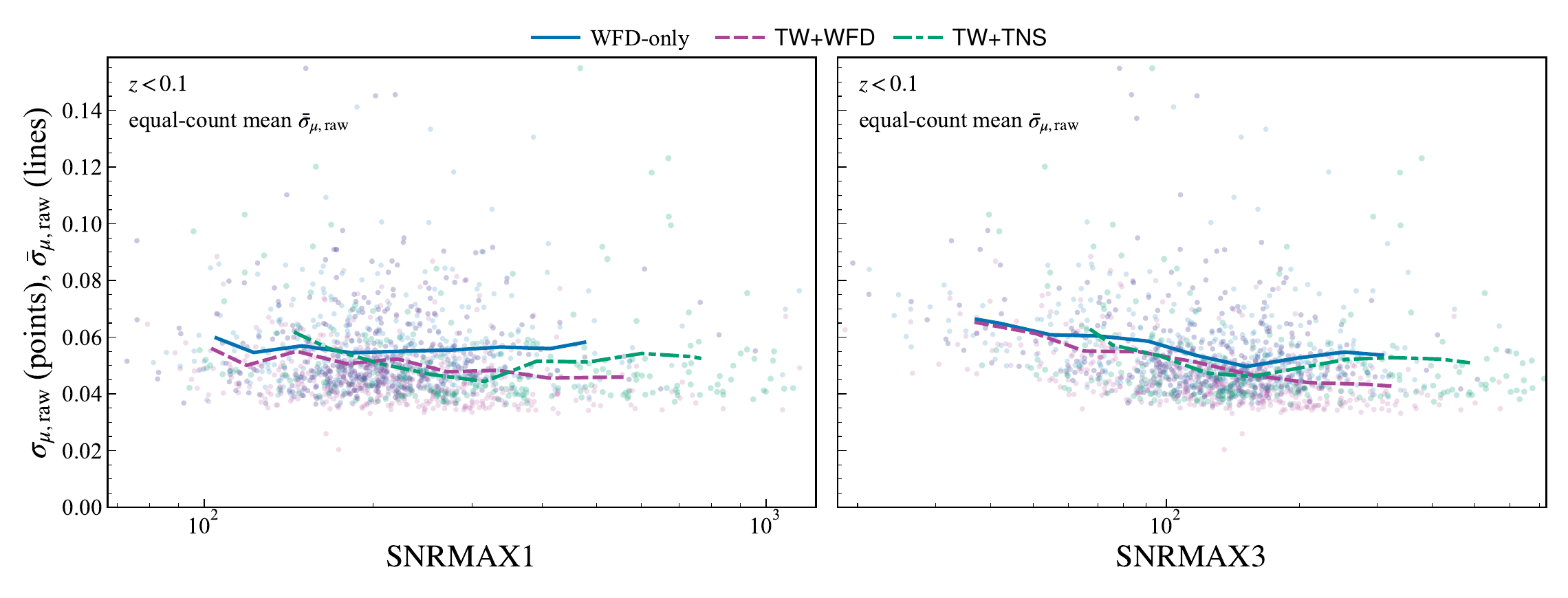}
\caption{\textbf{Measurement-only distance precision as a function of light-curve SNR.}
Faint points show the measurement-only distance-modulus uncertainty $\sigma_{\mu,\mathrm{raw}}$ in magnitudes (Equation~\ref{eq:sigmamu_raw}) for individual quality-selected SNe at $z<0.1$; curves show equal-count mean trends. The solid, dashed, and dash-dotted lines give \sampleWFD{}, \sampleTWWFD{}, and \sampleTWTNS{}, respectively. The vertical axis excludes intrinsic scatter, lensing, and peculiar-velocity noise. The two panels use the largest and third-largest values of the maximum single-epoch SNR among passbands used in each fit. At high SNR, $\bar{\sigma}_{\mu,\mathrm{raw}}$ flattens, so single-epoch depth is no longer the only limitation. For WFD-linked SNe, the \sampleTWWFD{} trend lies below the WFD baseline over most of the populated range, consistent with improved phase and band coverage. \sampleTWTNS{} represents an independent target population rather than matched WFD objects. \label{fig:snr_floor_twilight}}
\end{figure*}

Figure~\ref{fig:snr_floor_twilight} shows the measurement-only distance uncertainty from the SALT3 fit covariance:

\begin{equation}
\sigma_{\mu,\mathrm{raw}}^2 = \sigma_{m_B}^2+\alpha^2\sigma_{x_1}^2+\beta^2\sigma_c^2
+2\alpha\mathrm{C}_{m_Bx_1}-2\beta\mathrm{C}_{m_Bc}-2\alpha\beta\mathrm{C}_{x_1c},
\label{eq:sigmamu_raw}
\end{equation}

with $\alpha=0.14$ and $\beta=3.1$. This is the SALT3 fit-covariance contribution to Equation~\ref{eq:sigmamu_sn}; it excludes the added intrinsic-scatter, lensing, and peculiar-velocity terms. The mean trend flattens once the light curves reach high SNR. For the WFD-linked samples, the \sampleTWWFD{} trend lies below the WFD baseline over most of the populated SNR range, consistent with additional epochs reducing the fitted light-curve covariance. The comparison also includes changes in which events pass the quality cuts. Further precision gains would require improved modeling of intrinsic color, dust, calibration, and selection rather than additional photon statistics alone.

Beyond the statistical gains under a fixed intrinsic-scatter model quantified in our Fisher forecasts, a Rubin-system nearby sample could support future SALT3 retraining, dust and intrinsic-color studies, and calibration cross-checks \citep{lowz_shared_2508_10878,Anumba2026_SALT3training}.

Potential distance-ladder applications concern the rare SNe at $z<0.02$. These events are the most likely to saturate in 30\,s WFD visits near maximum. Losing multiband peak color and light-curve-shape information can compromise the standardized distance needed for precision $H_0$ work. TW observations reduce this risk by providing short-exposure light curves on the Rubin photometric system. Events in hosts with independent Cepheid or TRGB distances could serve as distance-ladder calibrators; other very nearby events can support nearest-Hubble-flow and local-velocity tests \citep{Riess2022SH0ES,Carrick2015_2Mpp,PantheonPlus_Redshifts,Kenworthy2022TwoRung}. Recent distance-ladder studies include a Coma-based distance calibration and a covariance-weighted Local Distance Network estimate \citep{scolnic_coma,H0DN}. A quantitative $H_0$ forecast would additionally require the expected number of SNe in calibrator hosts, host-distance uncertainties, a matched Hubble-flow comparison sample, and local-flow corrections.

\subsection{Implications for SN Modeling: Bands and Intrinsic Color/Dust}
\label{sec:modeling_implications}
Observing through the $z$ band in Rubin \emph{griz}---modeled here with SALT3.P22-NIR---provides redder optical measurements that can test fitted colors, $K$-corrections, and light-curve covariances in extended SALT3 models \citep{Kenworthy2021_SALT3,Pierel2022SALT3NIR,Taylor2023_SALT3update,Anumba2026_SALT3training}. Table~\ref{tab:snstats} shows that the effective per-SN distance uncertainty is nearly flat over $0.03\lesssim z\lesssim0.07$ under the adopted scatter prescription. The data could add Rubin-system objects to future intrinsic-color, dust-population, and calibration-training tests \citep{BroutScolnic2021_Dust,Popovic2023_Dust2Dust,Rubin2026_BananaSplit}. Reducing the intrinsic-scatter floor itself will require retraining and validating those models.

\subsection{Coordination with Rubin and Other Surveys}
\label{sec:ops_interplay}
The TW program would use candidate classifications, redshifts, recent photometry, and cadence information from Rubin brokers and partner surveys to identify where short exposures would fill gaps in coverage. High-priority cases include unsaturated peak-phase colors, missing bands, cadence gaps near peak, and very-low-redshift events needed for PV or distance-ladder analyses. Broker-selected alerts and evolving light curves could be retained in the LSST Dark Energy Science Collaboration's Fast Access to Survey Transients Database (FASTDB), a downstream database designed to ingest broker streams and preserve multiple processing versions of Rubin and DESC photometry \citep{LSSTDESC_FASTDB2025}. External photometry should be treated as trigger and validation information unless the cosmology analysis includes a full cross-calibration model.

The schedule contains roughly 50--90 full-focal-plane images per usable night. The raw data-transfer and processing load scales with image count. Before operations, representative 1, 2, and 5\,s images should be processed through instrument-signature removal, image differencing, source measurement, and alert production to measure artifact rates, photometric repeatability, and latency.

Calibration outside the main WFD footprint requires commissioning validation. The scheduler applies a nominal $35^\circ$ altitude floor during candidate feasibility checks. Because visits are subsequently ordered within each window, the observing plan extends to a minimum actual start altitude of $32.2^\circ$ ($X=1.87$); an operational implementation should check the minimum-altitude requirement again at the exposure start time. Rubin's calibration plan includes FGCM-style global calibration \citep{FagreliusRykoff2025_SITCOMTN086,Burke2017_FGCM}, but the uniformity achieved for 1--5\,s twilight exposures and non-WFD fields must be measured on sky. Detector nonlinearity should be tested across the signal range of these bright epochs. Rubin's per-amplifier correction is derived from flat-field pairs at different illumination levels monitored by a reference photodiode \citep[Section~3.9]{FagreliusRykoff2025_SITCOMTN086}. For TW observations, repeated stellar measurements in 1, 2, and 5\,s exposures should test the corrected flux ratios as a function of peak-pixel signal and sky background, including overlap with unsaturated 30\,s observations. These tests would establish the usable bright limit and residual calibration uncertainty. Northern or equatorial fields may also require PS1--LSST cross-calibration. In PV analyses, remaining large-angle calibration patterns should be tested with overlap regions and, when necessary, marginalized with low-order angular modes such as dipole or quadrupole terms.

ATLAS and ZTF are major contributors to the TNS classified-SN stream and produce substantial nearby-SN samples \citep{Tonry2018_ATLAS,Bellm2019_ZTF}. LS4, WFST, BlackGEM, GOTO-like surveys, DEBASS, and TiDES/4MOST may provide additional discovery, cadence, classification, or host-redshift information \citep{Dyer2022_GOTO,WFSTCollaboration2023,Groot2024_BlackGEM,Miller2025_LS4,lowz_shared_2508_10878,acevedo2025dark,Frohmaier2025_TiDES}. The forecast depends on the latency, classification reliability, and redshift completeness of these streams.

The forecast counts and statistical constraints assume reliable SN~Ia classifications and sufficiently accurate spectroscopic redshifts for every selected event. Classified TNS reports with redshifts supply the empirical trigger distribution; their reported redshifts do not establish uniform spectroscopic quality, and the completeness and timeliness of future follow-up require a separate selection model. Missing classifications or redshifts would reduce the usable sample and could introduce selection effects correlated with brightness, host properties, sky position, or redshift.

A targeted program on another facility, such as DECam, could provide complementary short-exposure follow-up. The number of SNe it could observe and their cosmological value would depend on its allocation, scheduler, filter set, saturation model, and cross-system calibration. The case evaluated here places the added measurements directly on the Rubin \emph{griz} system, allowing them to be merged with WFD epochs at the object level.
\subsection{Recommendation}
\label{sec:recommendation}
We recommend evaluating a regular Rubin twilight SN program with a hard $90^\circ$ solar exclusion, short \emph{griz} exposures, limited filter changes, and explicit accounting for dome-slew overheads. Under the assumptions in Section~\ref{sec:forecast-assumptions}, adding twilight reduces the fractional uncertainty on $f\sigma_8$ from 0.783 to 0.305 in $0.02\le z<0.06$ and increases the SN-only dark-energy FoM by 18.2\% in our one-year forecasts. The smaller gains in the $-12^\circ$ to $-11^\circ$ control show that the $-11^\circ$ to $-9^\circ$ interval should be adopted only after commissioning tests of sky brightness, photometric calibration, and visit execution.

\section*{Data and Code Availability}
The public repository for the twilight scheduler and associated analysis code is \url{https://github.com/trivialTZ/Lsst-Twilight-SNe}. The Rubin Data Preview~1 data used in this work \citep{Rubin_DP1_dataset_2025,RubinDP1_2026} are accessible to Rubin data-rights holders through the Rubin Science Platform, subject to the Rubin Observatory data access policy. The DP1 source selection used for the saturation analysis is described in Appendix~\ref{sec:sat_validation}.

\begin{acknowledgments}

This paper has undergone internal review in the LSST Dark Energy Science Collaboration.
We thank the internal reviewers, Michael Wood-Vasey and Richard Kessler, for their comments and suggestions.
We also thank Maria Vincenzi and Rujuta Purohit for their helpful comments on earlier drafts of the manuscript.

\textbf{Author contributions.} Xianzhe TZ Tang developed the twilight scheduler, adapted the analysis code, and wrote the manuscript. Dillon Brout contributed ideas, provided scientific guidance, and revised the manuscript. Christopher W. Stubbs contributed ideas and scientific guidance. Elana K. Urbach and J. Bryce Kalmbach provided advice on Rubin engineering constraints. Bastien Carreres provided cosmology code that was adapted for this study. Erik R. Peterson and Chris Frohmaier provided simulations of the Legacy Survey of Space and Time used in this study. Greg Aldering provided scientific guidance and expertise and contributed to manuscript revisions. Ayan Mitra contributed to the development of the simulation pipeline. Corentin Ravoux contributed to the development of \texttt{FLIP}.

TZ thanks the LSST-DA Data Science Fellowship Program, which is funded by LSST-DA, the Brinson Foundation, the WoodNext Foundation, and the Research Corporation for Science Advancement Foundation; his participation in the program has benefited this work. E.K.U. acknowledges support from the Vera C. Rubin Observatory through the National Science Foundation Cooperative Agreements AST-1258333 and AST-2241526 and Cooperative Support Agreements AST-1202910 and AST-2211468, managed by the Association of Universities for Research in Astronomy (AURA). The work of JBK was supported by DOE Contract DE-AC02-76SF00515. The work of G.A. was supported in part by the Director, Office of Science, Office of High Energy Physics of the U.S. Department of Energy under Contract No. DE-AC02-05CH11231.

The authors acknowledge that this research used resources of the National Energy Research Scientific Computing Center (NERSC), a U.S. Department of Energy Office of Science User Facility located at Lawrence Berkeley National Laboratory and operated under Contract No. DE-AC02-05CH11231; resources provided by the University of Chicago's Research Computing Center; and the Shared Computing Cluster, which is administered by Boston University's Research Computing Services.

The DESC acknowledges ongoing support from the Institut National de
Physique Nucl\'eaire et de Physique des Particules in France; the
Science \& Technology Facilities Council in the United Kingdom; and the
Department of Energy and the LSST Discovery Alliance
in the United States.  DESC uses resources of the IN2P3
Computing Center (CC-IN2P3--Lyon/Villeurbanne - France) funded by the
Centre National de la Recherche Scientifique; the National Energy
Research Scientific Computing Center, a DOE Office of Science User
Facility supported by the Office of Science of the U.S.\ Department of
Energy under Contract No.\ DE-AC02-05CH11231; STFC DiRAC HPC Facilities,
funded by UK BEIS National E-infrastructure capital grants; and the UK
particle physics grid, supported by the GridPP Collaboration.  This
work was performed in part under DOE Contract DE-AC02-76SF00515.

This material is based upon work supported in part by the National Science Foundation through Cooperative Agreements AST-1258333 and AST-2241526 and Cooperative Support Agreements AST-1202910 and 2211468 managed by the Association of Universities for Research in Astronomy (AURA), and the Department of Energy under Contract No. DE-AC02-76SF00515 with the SLAC National Accelerator Laboratory managed by Stanford University. Additional Rubin Observatory funding comes from private donations, grants to universities, and in-kind support from LSST-DA Institutional Members.

This publication is based in part on proprietary Rubin Observatory Legacy Survey of Space and Time (LSST) data, and was prepared in accordance with the Rubin Observatory data rights and access policies. All authors of this publication meet the requirements for co-authorship of proprietary LSST data.

This research uses services or data provided by the Rubin Science Platform at NSF-DOE Vera C. Rubin Observatory, which is jointly funded by the U.S. National Science Foundation and the U.S. Department of Energy, Office of Science.

\end{acknowledgments}

\facilities{Rubin:Simonyi (LSSTComCam), Rubin:USDAC}

\bibliographystyle{aasjournalv7}
\bibliography{sample701}

\appendix
\counterwithin{equation}{section}
\section{Scheduler Details}\label{app:scheduler}

This appendix documents the equations, configuration parameters, and output conventions underlying the four-stage scheduler summarized in Section~\ref{sec:scheduler-overview}. The planner produces visit tables and, for the simulations, a cadence library (SIMLIB) containing each planned exposure and its photometric metadata. The TNS and WFD input catalogs are described in Section~\ref{sec:inputs}.

\subsection{Survey Context and Twilight Windows}
\label{sec:twilight}
We use twilight observations with Rubin Observatory to recover bright, very-nearby SNe~Ia that are prone to being under-sampled (and in some cases saturated) in standard WFD visits, and to obtain unsaturated observations near peak. TNS reports supply the external low-$z$ trigger stream. The baseline WFD SIMLIB provides WFD targets and the timing of their scheduled visits. Objects present in both sources are paired by name and de-duplicated for planning. For the current WFD baseline relevant to our $gri$ saturation calculation, a standard visit is a 30~s exposure \citep{RubinKeyNumbers2026,RubinSurveyStrategyV5_2025}. Our twilight planner instead models single short exposures per filter, typically 1--5~s under the elevated solar twilight background, with a minimum inter-exposure spacing of 15~s enforced (Section~\ref{sec:filtering}). We schedule independently in the evening and morning \emph{twilight windows} of each local night, defined by Sun altitude $\mathrm{Alt}_\odot\in[\mathrm{Alt}_{\odot,\min},\mathrm{Alt}_{\odot,\max}]$; by default we adopt the darker half of nautical twilight, $-12^\circ\le\mathrm{Alt}_\odot\le-9^\circ$, with a customizable policy by Sun altitude for filter eligibility (Section~\ref{sec:policy}). We use Cerro Pach\'on site and telescope properties from \citet{ivezic2019lsst}.

We adopt the documented 120~s maximum filter-swap time and a nominal 2~s hardware readout \citep{RubinDMTN065}. The telescope slew model has an initial $3.5^\circ$ acceleration phase lasting 4~s, followed by a rate of $5.25^\circ$\,s$^{-1}$. We also require 15~s between exposure starts as a conservative planning constraint motivated by the potential short-exposure thermal limit in DMTN-065. All parameters are configurable. Figure~\ref{fig:twilight_overview} summarizes the planning flow.

\subsection{Solar Geometry, Sky-Model Domain, and Dome Movement}
\label{sec:sun_dome_policy}
\paragraph{Solar exclusion and directional twilight sky.}
The Sun and target are transformed to the topocentric frame at Cerro Pach\'on. A visit is feasible only if
\begin{equation}
\Delta_\odot(t)\ge90^\circ
\end{equation}
at the exposure start, midpoint, and end. This minimum separation is a fixed safety and operational requirement and does not vary with filter, solar phase, or solar altitude. It therefore differs from the graded Moon policy in Section~\ref{sec:policy}. The sky background remains direction dependent within the allowed hemisphere: the \emph{RubinSim} twilight model receives solar altitude, target airmass, and the target azimuth relative to the Sun, \texttt{azRelSun}, at each target's RA, Dec, and MJD \citep{RubinSkyBrightnessDocs2026}. Consequently, toward-Sun and anti-Sun candidates at the same airmass and solar altitude can have different $m_5$, SNR, and opportunity cost. Appendix~\ref{app:m5_zenith_sky} illustrates the resulting range.

The fitted \texttt{TwilightInterp} domain ends at $\mathrm{Alt}_\odot=-11^\circ$. We retain the $-12^\circ$ to $-9^\circ$ policy as the main operational experiment but flag every evaluation above $-11^\circ$ as extrapolated. The control policy, $-12^\circ\le\mathrm{Alt}_\odot\le-11^\circ$, remains within the fitted domain. Independent reconstruction of all main-policy visits gives a minimum $\Delta_\odot=90.003^\circ$ and no start-, midpoint-, or end-time violations; the control minimum is $90.008^\circ$. We do not add a separate terrain-horizon, enclosure-shadow, or dome-vignetting brightness correction. Such effects are distinct from the modeled solar azimuthal gradient and would require an observatory-specific horizon/stray-light map.

\paragraph{Dome kinematics.}
The telescope mount assembly (TMA) and enclosure axes can move simultaneously, so serialized execution requires
\begin{equation}
t_{\rm move}=\max\!\left(t_{\rm TMA},t_{\rm dome}\right).
\end{equation}
For strict recentering, the dome azimuth and elevation axes also move in parallel:
\begin{equation}
t_{\rm dome}=
\max\!\left[
t_{\rm jerk}\!\left(\Delta{\rm Az};v_{\rm Az},a_{\rm Az},j_{\rm Az}\right)
+1\,{\rm s}\,\mathbb{I}_{\Delta{\rm Az}>0},
t_{\rm jerk}\!\left(\Delta{\rm El};v_{\rm El},a_{\rm El},j_{\rm El}\right)
\right].
\end{equation}
We use the Rubin design-minimum values $(v,a,j)_{\rm Az}=(1.50~{\rm deg\,s^{-1}},0.75~{\rm deg\,s^{-2}},3.00~{\rm deg\,s^{-3}})$ and $(v,a,j)_{\rm El}=(1.75~{\rm deg\,s^{-1}},0.875~{\rm deg\,s^{-2}},3.50~{\rm deg\,s^{-3}})$, together with the Rubin scheduler's jerk-limited motion profile \citep{RubinTelescopes2026}. These are nominal design requirements, not final measured on-sky performance. The first target in each window is treated as prepositioned. We do not credit predictive enclosure crawling during an exposure, making this a conservative strict-recenter model. Reconstructing every inter-target move from the visit table reproduces the stored dome angles and times to the output rounding precision. The dome limits 85.8\% of main-policy moves and 84.5\% of control moves.

\paragraph{Forecast route ranking.}
In the one-year forecast schedules, the movement term for ranking feasible candidates uses a TMA angular-distance proxy. Feasibility and window-capacity calculations include the full telescope--dome movement time $t_{\rm move}$.

\subsection{Inputs, Ephemerides, and Sky/Noise Model}
\label{sec:inputs}
\paragraph{Inputs.}
The planner ingests: (i) a \emph{planner configuration} (Section~\ref{sec:config}); (ii) a \emph{sky model configuration} (airmass, sky brightness, seeing/FWHM, zeropoints); (iii) a \emph{candidate catalog} built by combining public, CSV-formatted TNS search results (object name, J2000 RA/Dec, discovery epoch, type, redshift when available, and discovery photometry) with a catalog-like table derived from a baseline WFD SIMLIB, then de-duplicating by IAU name (TNS metadata retained where available); (iv) an optional \emph{external cadence map} from the same WFD SIMLIB, keyed by object name and filter, providing scheduled WFD visit times used to check the minimum spacing between visits and calculate the cadence bonus; (v) per-type typical \emph{lifetime-days} and visibility window; and (vi) a streaming \emph{history} of past TW observations used for cadence/priority. For SIMLIB-derived candidates that lack a discovery date, we approximate the discovery epoch with the first WFD visit time in the SIMLIB.

\paragraph{Ephemerides and airmass.}
We precompute the Sun/Moon altitude and target alt/az across each twilight window on a grid (default 2~min). Airmass $X$ as a function of zenith angle $\gamma$ uses the Kasten--Young approximation \citep{kasten1989airmass}:
\begin{equation}
\label{eq:ky}
X(\gamma) = \left[\cos\gamma + 0.50572\,(96.07995^\circ - \gamma)^{-1.6364}\right]^{-1}.
\end{equation}
This approximation is valid for $\gamma<90^\circ$.

\paragraph{Sky brightness and Moon.}
The sky model combines a dark-sky baseline with a solar twilight component evaluated from solar altitude, target airmass, and target azimuth relative to the Sun, plus a lunar component. The same direction-dependent background enters feasibility, $m_5$, and the opportunity-cost term used for visit ordering. For the Moon we adopt the \citet{krisciunas1991moon} scattering prescription as a function of phase, target--Moon separation, and Moon/target airmass. In the bright twilight regime considered here, the lunar term is often subdominant to the solar twilight background but is retained for conservative filter and depth decisions. A Sun-altitude policy (Section~\ref{sec:policy}) restricts usable filters as twilight brightens.

\paragraph{Photometric depth and SNR.}
We follow the LSST single-visit depth formalism of \citet[their Equations~(6) and (7)]{ivezic2019lsst} for the $5\sigma$ single-visit depth $m_5$ and its dependence on sky surface brightness $\mu_{\rm sky}$ ($\mathrm{mag}\,\mathrm{arcsec}^{-2}$), effective seeing $\Theta$ (arcsec), exposure time $t_{\rm vis}$, and airmass $X$. For the Opsim-style implementation used here, we adopt the algebraically equivalent form summarized in SMTN-002 \citep{jones2017smtn002}\footnote{\url{https://smtn-002.lsst.io/}; see the section ``Calculating $m_5$ values in the LSST Operations Simulator.''}. Equation~\ref{eq:m5} is written in terms of a band-dependent reference depth $m_{5,\mathrm{ref}} \equiv C_m + 0.50\,(\mu_{\mathrm{dark}}-21)$, where $C_m$ is the throughput constant in the SMTN-002 notation. We use the standard SNR model
\begin{equation}
\label{eq:snr}
\mathrm{SNR} = \dfrac{C}{\sqrt{C + n_{\mathrm{eff}}\,\left(B + \sigma_{\mathrm{RN}}^2\right)}}.
\end{equation}
Here $C$ is source electrons, $B$ is sky electrons per pixel, $n_{\mathrm{eff}}$ is the noise-equivalent area in pixels (from the PSF), and $\sigma_{\mathrm{RN}}$ is read noise ($\mathrm{e}^-$). The corresponding 5$\sigma$ depth can be written as
\begin{equation}
\label{eq:m5}
m_5 = m_{5,\mathrm{ref}}
\; +\; 0.50\,\big(\mu_{\mathrm{sky}}-\mu_{\mathrm{dark}}\big)
\; +\; 2.5\,\log_{10}\!\left(\dfrac{0.7''}{\Theta}\right)
\; +\; 1.25\,\log_{10}\!\left(\frac{t_{\mathrm{vis}}}{30\,\mathrm{s}}\right)
\; -\; k_F\,(X-1)
\; +\; \Delta C_m(\tau).
\end{equation}
Here $k_F$ is the extinction coefficient for filter $F$, and $\Delta C_m(\tau)$ is the short-exposure/read-noise correction, corresponding to $dC_m(Tscale)$ in SMTN-002 and to Equation~(7) of \citet{ivezic2019lsst}, with $\tau = (t_{\mathrm{vis}}/30\,\mathrm{s})\,10^{-0.4(\mu_{\mathrm{sky}}-\mu_{\mathrm{dark}})}$. This correction accounts for the departure from simple sky-limited depth scaling when short exposures make read noise non-negligible. We require feasibility $\mathrm{SNR}\ge\mathrm{SNR}_{\min}$ (default 5) or, equivalently, that the target be brighter than the 5$\sigma$ depth by $\ge0$~mag, i.e., $m_5 - m_{\mathrm{target}} \ge 0$.
We do not include a separate flat-field noise floor in Eq.~\ref{eq:snr}; under a high solar-twilight background such a term would scale approximately as $\sigma_{\rm ff} B$ and should be included in an operations-level photometric error budget.

\subsection{Candidate Filtering}
\label{sec:filtering}
We apply the following selection criteria before assigning priorities:
\begin{itemize}
\item \textbf{Altitude} $h\ge h_{\min}$ (we use $35^\circ$) at some time within the window.
\item \textbf{Solar policy}: require topocentric $\Delta_\odot\ge90^\circ$ at exposure start, midpoint, and end, and allow only filters permitted by $\mathrm{Alt}_\odot$ (Section~\ref{sec:policy}).
\item \textbf{Moon separation} $\Delta{\rm Moon}\ge D_F$ per filter $F$ (e.g., $D_g\simeq50^\circ$, $D_r\simeq35^\circ$, $D_i\simeq30^\circ$, $D_z\simeq25^\circ$, $D_y\simeq20^\circ$), relaxed when the Moon is faint or low using the altitude/phase weight defined in Section~\ref{sec:policy}.
\item \textbf{Minimum SNR}: Eq.~\eqref{eq:m5} predicts ${\rm SNR}\ge{\rm SNR}_{\min}$ using the sky/throughput model at a representative time in the window (or, when enabled, at the time of best expected conditions within that window), i.e., require $m_5 - m_{\mathrm{target}} \ge 0$.
\item \textbf{Saturation check}: predict peak-pixel electrons from the source counts and PSF; require peak $< N_{\rm sat}$ (default $8\times10^4$~$\mathrm{e}^-$). When adaptive exposure capping is enabled, if violated, reduce exposure time $t$ until
\begin{equation}
\label{eq:msat}
m_{\mathrm{sat}}(t) = m_{\mathrm{sat}}(t_0) + 2.5\,\log_{10}\!\left(\dfrac{t}{t_0}\right)
\end{equation}
falls below the target's magnitude, or until the configured floor (1--5~s). For example, shortening a 30~s exposure to 5~s gives $2.5\log_{10}(5/30)=-1.95$~mag in this sign convention, so the observation can tolerate sources about 1.95~mag brighter before reaching the same peak-pixel charge. When this cap applies, we recompute feasibility and expected SNR at the reduced exposure using the same photometric model. This adaptive cap is a forecast-level saturation-avoidance rule: the simulated scheduler has target magnitudes from the input catalog and light-curve model, whereas a real-time implementation would need to use only pre-observation information such as broker photometry, recent Rubin measurements, host association or redshift when available, and a conservative safety margin. A fixed short-exposure mode can be evaluated with the same machinery by disabling the adaptive cap. In the one-year main plan, 96\% of scheduled visits use the full 5\,s exposure and the cap shortens only the remaining ${\sim}4\%$, so the forecast is already close to a fixed 5\,s program.
By default we use \texttt{SNANA}'s Taylor approximation for the central-pixel fraction $f_A$ when evaluating the peak-pixel charge (exact Gaussian integration is also supported); this ensures consistency with our SIMLIB export.

\item \textbf{Additional selection criteria}: optional limits on target age and time since discovery, and a minimum interval between same-filter visits. With the external WFD cadence map, the scheduler checks the nearest preceding and following visits in the same band. A TW visit is rejected if either WFD visit falls within the exclusion interval, even if that 30\,s WFD exposure would saturate. This limitation can reduce recovery of near-peak photometry for the brightest events.

\end{itemize}

\subsection{Priority Scoring}
\label{sec:priority}
Each feasible SN--filter pair $(i,F)$ receives a scalar score following the implementation:
\begin{equation}
\label{eq:prio}
\Pi{(i,F)} \;=\; \underbrace{\mathrm{Need}_i}_{\in\{0,1\}}
\times \underbrace{\mathrm{LowZBoost}_i}_{\ge 1}
\times \left[1 + \underbrace{B_{{\rm cad},i}(F)}_{\text{Gaussian cadence bonus}}\right]
\times \left[1 + \varepsilon_{\rm snr}\,\max\!\big(0,\; m_{5,i}(F)-m_i(F)\big)\right].
\end{equation}
Here $\mathrm{Need}_i$ comes from the priority tracker (1 if the target still needs observations under the configured strategy, else 0). The factor $\mathrm{LowZBoost}_i\ge1$ upweights low-$z$ SNe~Ia (and may include a smooth redshift boost). The cadence bonus $B_{{\rm cad},i}(F)$ is a Gaussian in an effective cadence offset $\Delta t_{\rm eff}=\min(t_{\rm now}-t_{\rm prev},\,t_{\rm next}-t_{\rm now})$, where $t_{\rm prev}$ is the most recent visit in filter $F$ and $t_{\rm next}$ the next scheduled visit (from WFD when available); $t_{\rm prev}$ and $t_{\rm next}$ are evaluated on the union of past TW visits and the external cadence map, and include first-epoch and color/diversity terms.
The optional SNR-margin term multiplies the score by $1+\varepsilon_{\rm snr}\max(0,m_5-m_{\rm target})$, so positive $\varepsilon_{\rm snr}$ favors candidates predicted to be farther above the SNR=5 threshold. In the reported forecast $\varepsilon_{\rm snr}=0$; the factor has no effect.
 The preferred initial filter is $\arg\max_F \Pi_i(F)$ with deterministic tie-breakers. The pair score is then passed to the filter-selection step described in Section~\ref{sec:dp}.

\subsection{Filter Policy and Conditions for Filter Changes}
\label{sec:policy}
We allow a configurable Sun-altitude policy reflecting twilight practicality (customizable by the user); for example, one can set: at $-18^\circ < \mathrm{Alt}_\odot \le -15^\circ$, prefer $y,z,i$. By default we evaluate eligibility at the time within the window when the target is predicted to have the best observing conditions (a midpoint evaluation is available as a fallback). Filter changes incur an explicit penalty equal to the \emph{filter-change time} $t_{\mathrm{swap}}\simeq120$~s \citep{bianco2022rubin}. We impose a per-window swap cap $S$ (default $S{=}2$) and require any filter change to clear a minimum-gain condition: the predicted increase in science value must exceed a specified threshold and enough time must remain for that gain to offset the lost observing time from the swap. We cap distinct filters per visit to a configured limit (default 1); when $>1$ is allowed, a complementary color band may be added when feasible.

In the tests shown here, the first filter follows a simple cycle that rotates by window to maintain band diversity (morning: $g,r,i,z$; evening: $i,z,g,r$). Later filter choices are still determined by the dynamic-programming step if a different sequence yields a higher total value.

\paragraph{Moon constraint.}
{
Let $\Delta_{\rm Moon}$ be the target--Moon angular separation, $F$ the Rubin filter, $s_F^{\rm base}$ the baseline minimum separation for that filter, $h_{\rm Moon}$ the Moon altitude in degrees, and $\Phi_{\rm Moon}$ the illuminated lunar fraction. In our implementation, $\Phi_{\rm Moon}=0.5[1-\cos(\psi_{\odot{\rm M}})]$, where $\psi_{\odot{\rm M}}$ is the Sun--Moon angular separation. We require
\[
\Delta_{\rm Moon} \ge s_F^{\rm eff}(h_{\rm Moon},\Phi_{\rm Moon}),
\qquad
s_F^{\rm eff}=s_F^{\rm base}\,W(h_{\rm Moon},\Phi_{\rm Moon}),
\]
where $s_F^{\rm base}=\{50^\circ,35^\circ,30^\circ,25^\circ,20^\circ\}$ for $F=\{g,r,i,z,y\}$ in the fiducial configuration. The dimensionless weight is
\[
W(h_{\rm Moon},\Phi_{\rm Moon}) =
\begin{cases}
0, & h_{\rm Moon}\le -10^\circ,\\
\dfrac{h_{\rm Moon}+10^\circ}{10^\circ}\,
\mathrm{clip}\!\left(0.3+0.7\Phi_{\rm Moon},\,0.3,\,1.0\right), & -10^\circ<h_{\rm Moon}<0^\circ,\\
1, & h_{\rm Moon}\ge0^\circ,
\end{cases}
\]
with $\mathrm{clip}(x,0.3,1.0)$ limiting $x$ to the interval $[0.3,1.0]$. We impose no minimum Moon--target separation when the Moon is more than $10^\circ$ below the horizon and restore the baseline separation once the Moon is above the horizon.
}

\subsection{Dynamic-Programming Filter Scheduling}
\label{sec:dp}
Within a window of length $T_{\rm win}$, let the current carousel filter be $f_s$. We consider at most $S$ swaps, which define $S{+}1$ filter \emph{segments}. For each feasible pair $(\mathrm{SN},F)$ we compute a \emph{pair score} $v(\mathrm{SN},F)$ by combining Eq.~\eqref{eq:prio} with feasibility margins (SNR/$m_5$) and any saturation risk penalty. Let ${(\mathrm{SN}_i,F)}$ be the filter-ordered list (per $F$) sorted by density $v/t_{\rm visit}$, and define the filter prefix value $P_F[k]=\sum_{j=1}^{k} v(\mathrm{SN}_j,F)$ with a corresponding time prefix $T_F[k]$.

We optimize the total value $\sum_j v(\mathrm{SN}_j,F_j)$ subject to a time budget and swap penalties by a DP over segments. One convenient recurrence indexed by the number of segments $s$ and the number of visits (or normalized ``slots'') $n$ is
\begin{equation}
\label{eq:dp}
V^{(s)}(n) =
\max\!\left(
 V^{(s)}(n-1) + \Delta P_{f_s}(1),\
 \max_{f_{s+1}\neq f_s}\left[ V^{(s+1)}(n-1) + \Delta P_{f_{s+1}}(1) - \lambda_{\mathrm{swap}}\right]
\right),
\end{equation}
with boundary $V^{(0)}(0)=0$, a swap budget $s\le S$, and $\lambda_{\mathrm{swap}}\equiv t_{\rm swap}/t_{\rm ref}$ the swap cost in visit-equivalents for a reference visit time $t_{\rm ref}$ (guard$+$exposure$+$readout$+$typical slew). Here $\Delta P_f(k)$ denotes the marginal increase in payoff from adding the $k$-th visit in filter $f$. In practice we solve a visit-budget variant using prefixes: for a candidate segment sequence $(f_1,\ldots,f_m)$ with $m\le S+1$, the optimal per-segment counts $(k_1,\ldots,k_m)$ maximize
\begin{equation}
\label{eq:objective}
\mathcal{V} = \sum_{j=1}^{m} \alpha_j\,P_{f_j}(k_j) - (m-1)\,\Lambda_{\mathrm{swap}}\quad\text{s.t.}\quad
\sum_{j=1}^{m} T_{f_j}(k_j) + (m-1)\,t_{\rm swap} \le T_{\rm win},
\end{equation}
where a small additional penalty after each swap discourages rapid alternation between filters. We also require a multi-filter schedule to improve on the best single-filter solution by at least a specified fractional amount before adopting it. In practice we evaluate only short sequences (small $m$), solve for the corresponding $(k_j)$, and retain a multi-filter schedule only when that improvement criterion is met. The DP output is therefore an ordered sequence of filter intervals and the number of visits assigned to each one.

\subsection{Visit Ordering within a Window}
\label{sec:greedy}
Given the DP-selected filter intervals, we order visits by minimizing a timing cost in seconds that compares the observing conditions available \emph{now} with the best conditions available for that target later in the same window and includes a small additive movement term:
\begin{equation}
\label{eq:cost}
C = t_{\rm vis}\,\big(10^{0.8\,\Delta m_5} - 1\big) + w_{\mathrm{move}}\,t_{\mathrm{route}} + \mathbb{I}_{\rm swap}\,\Pi_{\rm swap}.
\end{equation}
Here $t_{\rm vis}$ is a conservative per-visit wall time, $\Delta m_5 = \max\big(0, m_{5,\,\rm best}-m_{5,\,\rm now}\big)$ compares the $5\sigma$ depth at the target's best time in the window to \emph{now} (both evaluated at the target's RA/Dec using the direction-dependent sky model in Eq.~\ref{eq:m5}), $w_{\rm move}$ is a small weight, $t_{\rm route}=t_{\rm move}$ in the current code, and $\Pi_{\rm swap}$ is a time penalty for a carousel filter swap when applicable. We enforce the twilight budget and per-target caps as hard constraints (including inter-exposure guards); candidates that would overflow $T_{\rm win}$ after accounting for exposure, readout, guard, telescope/dome movement, and filter-change times are skipped. Filter use within a window is set by the dynamic-programming calculation described above, which tests a small number of ordered filter blocks while including the $\sim120$~s carousel swap penalty and a modest extra penalty for closely spaced swaps. Within each block, targets are ranked by priority per unit observing time, subject to the minimum visit intervals and per-SN caps. If time remains after the planned blocks are filled, the scheduler can add repeat observations or previously deferred targets, subject to the same constraints. As noted in Section~\ref{sec:sun_dome_policy}, the one-year forecast uses a TMA angular-distance proxy for route ranking while enforcing full telescope--dome movement times in feasibility and budget checks.

\subsection{Outputs and \texttt{SNANA} Interface}
\label{sec:outputs}
\paragraph{Per-window tables.}
For each window, the scheduler outputs a summary (candidates, selected visits, swaps used, utilization, median airmass/sky, and cadence statistics) and a time-ordered visit table with a full timing breakdown (slew, readout, guard, exposure, and filter changes). Altitude, airmass, sky surface brightness, and zeropoints are evaluated at the \emph{actual visit start time} for each row (i.e., not at the time of best conditions within the window).

\paragraph{\texttt{SNANA} SIMLIB export.}

For simulation, the planner also produces an observing-cadence library (SIMLIB) for \texttt{SNANA} \citep{kessler2009snana}. The planner-native library is subsequently split or merged with the input WFD library; conversion preserves the physical twilight PSF width while expressing it in the destination library's declared units.

The planner-native library contains a global header
 (\texttt{SURVEY}, \texttt{FILTERS}, \texttt{PIXSIZE}, \texttt{NPE\_PIXEL\_SATURATE}, \texttt{PHOTFLAG\_SATURATE}, \texttt{PSF\_UNIT})
and one \texttt{LIBID} block per SN. The split and merged SIMLIBs used for the sample-count and dark-energy fits omit the two saturation keys, whereas the saturation-enabled PV runs set \texttt{NPE\_PIXEL\_SATURATE=80000} and \texttt{PHOTFLAG\_SATURATE=2048} in the simulation inputs and \texttt{PHOTFLAG\_MSKREJ=2048,0} in the fit inputs to reject flagged epochs (Section~\ref{sec:snana}).
 Each block has \texttt{EPOCH} (\texttt{S:}) lines of the form: MJD, \texttt{ID+NEXPOSE}, \texttt{FLT}, \texttt{GAIN}, \texttt{NOISE}, \texttt{SKYSIG}, \texttt{PSF1}, \texttt{PSF2}, \texttt{PSFRATIO}, \texttt{ZPTAVG}, \texttt{ZPTERR}, where:
\begin{itemize}
\item \textbf{Timing and filter.} MJD is the actual visit start time; \texttt{FLT} is the band (with y written as uppercase \texttt{Y} per \texttt{SNANA} convention). When an epoch consists of multiple equal snaps, \texttt{ID+NEXPOSE} is formatted as \texttt{k*NEXPOSE}; otherwise \texttt{ID} alone is used.
\item \textbf{Zeropoint.} We compute the electron zeropoint
\(ZPT_{\rm pe} = ZPT_{1\,\mathrm{s}} + 2.5\,\log_{10} t_{\rm exp} - k_F\,(X-1)\),
using the per-band 1\,s zeropoint, the actual exposure time $t_{\rm exp}$ (after any saturation capping), extinction $k_F$, and the Kasten--Young airmass $X$ \citep{kasten1989airmass}. The ADU zeropoint is \(ZPTAVG = ZPT_{\rm pe} - 2.5\,\log_{10} \mathrm{GAIN}\). \texttt{ZPTERR} is a small configurable zeropoint uncertainty.

\item \textbf{Sky noise.} The sky surface brightness $\mu_{\rm sky}$ is evaluated at (RA,Dec,MJD) using \emph{RubinSim} when available; otherwise we use a Sun-altitude twilight term plus a Krisciunas\,\&\,Schaefer lunar model. We convert $\mu_{\rm sky}$ to per-pixel electrons via
\(B_{\rm pix}\,[\mathrm{e}^-] = 10^{-0.4(\mu_{\rm sky} - ZPT_{\rm pe})}\,\times\,\mathrm{PIXSIZE}^2\).
In the planner-native library, \texttt{SKYSIG} is $\sqrt{B_{\rm pix}}/\mathrm{GAIN}$ in ADU per pixel; \texttt{NOISE} is read noise in electrons per pixel. Split and merged libraries convert \texttt{SKYSIG} to the destination unit declared by \texttt{SKYSIG\_UNIT}.

\item \textbf{PSF parameters.} We adopt a Gaussian-equivalent effective FWHM by band. For pixel scale $p$, its Gaussian width is $\sigma_{\rm pix}=\mathrm{FWHM}/(2.355p)$ and its noise-equivalent area is $\mathrm{NEA}=2.266(\mathrm{FWHM}/p)^2$ pixels$^2$. The planner-native library declares \texttt{PSF\_UNIT=PIXEL\_SIGMA} and writes this $\sigma_{\rm pix}$ as \texttt{PSF1}, with \texttt{PSF2=PSFRATIO=0}; it does not write NEA separately. The simulation libraries derived from it declare \texttt{PSF\_UNIT=ARCSEC\_FWHM} and express twilight PSF1 as FWHM in arcseconds. For example, the $i$-band twilight value is 1.762 pixels in the native library and 0.830\arcsec{} in the simulation input.

\end{itemize}
We do not include an explicit airmass column in the \texttt{SNANA} SIMLIB; $X$ is already encoded through \texttt{ZPTAVG} and \texttt{SKYSIG}. All per-epoch quantities are evaluated at the \emph{actual visit start time} using the same sky and throughput model as for $m_5$ (Eq.~\ref{eq:m5}) and the SNR definition (Eq.~\ref{eq:snr}),
keeping the planner-generated twilight noise inputs aligned with the depth model.
 Unless overridden, we adopt \texttt{GAIN}$=1.6$\,e$^{-}$/ADU, \texttt{PIXSIZE}$=0.200$\,arcsec, and a band-dependent read noise.

\paragraph{WFD-aware split/merge products.}
When a baseline WFD SIMLIB is available, the pipeline provides two post-processing steps for end-to-end simulations: (i) split TW epochs into TNS-only and WFD-linked SIMLIBs (for attribution), and (ii) merge the TW epochs into the baseline WFD SIMLIB, preserving \texttt{LIBID}s,
to create a single cadence library for the matched WFD events with appended TW observations (the \sampleTWWFD{} case).

\subsection{Configuration, Reproducibility, and Defaults}
\label{sec:config}
Key configuration parameters (defaults in brackets):
\begin{itemize}
\item \textbf{Twilight bounds:} $\mathrm{Alt}_{\odot,\min},\mathrm{Alt}_{\odot,\max}$ [$-12^\circ,-9^\circ$ main; $-12^\circ,-11^\circ$ calibrated control]; hard $\Delta_\odot\ge90^\circ$ at start, midpoint, and end; window step 2~min.
\item \textbf{Visibility:} $h_{\min}$ [$35^\circ$ for candidate feasibility in this work; code default $30^\circ$; the planned visit sequence reaches $32.2^\circ$ at actual start time, motivating an execution-time recheck]; Moon separations $D_F$ [$80^\circ,50^\circ,35^\circ,30^\circ,25^\circ,20^\circ$ for $ugrizy$].
\item \textbf{Exposure \& visit:} per-filter exposures configurable (typ. 1--5~s in twilight); the saturation check can cap exposures to avoid central-pixel full well (charge capacity); readout 2~s; minimum inter-exposure spacing 15~s.
\item \textbf{Overheads:} filter swap $t_{\rm swap}\approx120$~s; carousel capacity 5; TMA slews follow a piecewise model (first $3.5^{\circ}$ in 4~s, then $5.25^{\circ}\,\mathrm{s}^{-1}$) with $\sim$1~s settle; dome axes use the strict jerk-limited model in Section~\ref{sec:sun_dome_policy}; total movement time is $\max(t_{\rm TMA},t_{\rm dome})$; inter-exposure guard 15~s \citep{bianco2022rubin,RubinTelescopes2026}.
\item \textbf{Photometry:} $m_5$ per Eq.~\ref{eq:m5} using the LSST depth formalism \citep{ivezic2019lsst}; airmass Eq.~\ref{eq:ky} \citep{kasten1989airmass}; sky brightness from \emph{RubinSim} at (RA,Dec,MJD) when available, else a Krisciunas\,\&\,Schaefer-style Sun\,+\,Moon model evaluated at the target's position and time \citep{krisciunas1991moon}. We adopt a gain of 1.6 e$^{-}$/ADU for conversions. When per-band source magnitudes are unavailable, per-filter brightness is estimated from a conservative combination of a catalog discovery magnitude (with color priors) and redshift via distance modulus and peak standardization; the brighter estimate per filter is used to ensure a safe saturation cap.
\item \textbf{Selection criteria:} $\mathrm{SNR}_{\min}{=}5$; central-pixel saturation cap $N_{\rm sat}{\sim}8\times10^4$~$\mathrm{e}^-$; target-age and SN-type visibility windows.
\item \textbf{Priority:} quick-color need flag; cadence target per filter (e.g., 3~d with a Gaussian cadence bonus centered on the target revisit interval); low-$z$ Ia boost; optional small SNR-margin term (Eq.~\ref{eq:prio}). A per-filter band-diversity weighting is enabled by default to favor under-observed bands; this policy is configurable.
\item \textbf{Filter planning:} explicit per-swap cost; optional minimum gain required before changing filters; optional cap on swaps per window; and visit budgeting within each window. In our tests the first filter follows a simple cycle by window to preserve diversity (morning: $g,r,i,z$; evening: $i,z,g,r$), while later filter choices are set by the dynamic-programming step if a different sequence yields a higher total value.
\item \textbf{Execution:} within the selected filter intervals, targets are ranked by priority per unit visit time with direction-dependent sky opportunity cost and effective TMA/dome movement; minimum visit intervals and per-SN caps are enforced, and repeat observations can be added if time remains.
\item \textbf{Outputs:} per-SN plan CSV, time-ordered visit CSV, per-window summary CSV, and optional SIMLIB with consistent photometric metadata \citep{kessler2009snana}.
\end{itemize}
The scheduler output is deterministic for fixed inputs and configuration; random seeds affect only optional tie-breaks when enabled.

Appendices~\ref{sec:psf_validation} and~\ref{sec:sat_validation} validate the PSF/NEA and saturation-threshold assumptions used above.

{
\section{PSF Model and NEA Validation}
\label{sec:psf_validation}

Here we test if departures from the Gaussian PSF assumed for scheduler SNR, depth, and central-pixel calculations materially change predicted saturation rates. The scheduler and native cadence library use an effective Gaussian FWHM: its Gaussian-computed noise-equivalent area (NEA) represents the assumed PSF-fitting noise area, and the same width sets the Gaussian central-pixel fraction used for the saturation check. The SIMLIB records PSF width, not a separate NEA field (Section~\ref{sec:outputs}). The final \texttt{SNANA} simulation inputs express that width as arcsecond FWHM after unit conversion. Table~\ref{tab:psf_params} gives the widths and corresponding Gaussian NEAs for a 0.2\arcsec/pixel scale. The Moffat comparisons below test sensitivity to wings that a single effective Gaussian width cannot encode. These are separate saturation-validation simulations. The sample counts and dark-energy results use fits without saturation rejection, whereas the PV forecast uses the saturation-rejected fits described in Section~\ref{sec:snana}.

\begin{table}[h]
\centering
\caption{Effective PSF Parameters for SNR and Saturation Calculations}
\label{tab:psf_params}
\begin{tabular}{lcccc}
\hline
Band & FWHM ($\Theta$) & $\sigma_{\rm pix}$ (PSF1) & NEA (pix$^2$) & NEA (arcsec$^2$) \\
\hline
$u$ & 0.90\arcsec & 1.911 & 45.89 & 1.836 \\
$g$ & 0.85\arcsec & 1.805 & 40.93 & 1.637 \\
$r$ & 0.83\arcsec & 1.762 & 39.03 & 1.561 \\
$i$ & 0.83\arcsec & 1.762 & 39.03 & 1.561 \\
$z$ & 0.85\arcsec & 1.805 & 40.93 & 1.637 \\
$y$ & 0.90\arcsec & 1.911 & 45.89 & 1.836 \\
\hline
\end{tabular}
\end{table}

\paragraph{Moffat PSF impact on saturation.}
Atmospheric PSFs can exhibit extended wings that are often approximated by a Moffat profile with power-law index $\beta \approx 2$--5. At fixed FWHM this increases the NEA relative to a Gaussian. For a Moffat profile with $\beta = 2.033$ \citep{trujillo2001}, the NEA ratio relative to Gaussian is 2.451, corresponding to a sky-limited depth penalty of $\Delta m_5 \approx 0.49$~mag.

We quantify the impact on saturation using \texttt{SNANA} simulations comparing Gaussian and Moffat PSF models with 1,200 SNe per configuration. In these fixed-FWHM tests the epoch-level saturation fraction changes by at most 6.3\% (relative), with the $r$ band showing the largest reduction. This behavior is expected because Moffat wings redistribute flux over more pixels and reduce the central-pixel charge even though the formal NEA is larger.

For a realistic WFD comparison we also use a Moffat model representative of Rubin seeing (Moffat $\beta = 4.765$, corresponding to a Kolmogorov atmosphere, with mean FWHM $= 0.86\arcsec$). The resulting WFD epoch-level saturation fractions are 2.74\% in $g$, 3.50\% in $r$, 2.09\% in $i$, and 0.02\% in $z$, compared with the baseline Gaussian estimates of 2.72\%, 3.43\%, 2.05\%, and 0.02\%, respectively. The near-cancellation reflects two competing dependencies: improved seeing concentrates flux and increases the peak-pixel charge, while extended wings redistribute flux and decrease it.

\paragraph{PSF non-uniformity in short exposures.}
Short twilight exposures (1--5~s) may experience PSF spatial variations that degrade PSF-fitting photometry, effectively increasing the noise-equivalent area by a factor $f$. This yields a depth loss $\Delta m_5 = 1.25 \log_{10}(f)$. These tests approximate the effect as an NEA increase; they do not model possible PSF goodness-of-fit failures, which should be checked with on-sky short-exposure data.

We tested NEA inflation factors $f = 1.5$ ($\Delta m_5 = 0.22$~mag) and $f = 2.0$ ($\Delta m_5 = 0.38$~mag) using \texttt{SNANA} simulations with 1,200 SNe per configuration. In these tests, NEA inflation increases the estimated saturation fraction by 19--44\% in $r$ band for $f = 1.5$--2.0. The increase is a selection effect: reduced depth leaves a shallower detected sample, so a larger fraction of the retained SNe lie close to the saturation boundary. For $f \approx 1.5$, the increase is $\sim$20\%.

We use these inflation tests to bound the sensitivity of the saturation estimates to PSF degradation in short exposures. The baseline rates (2--3\%) represent ideal PSF-fitting conditions; accounting for spatial non-uniformity yields 2.4--4.2\% ($f = 1.5$) or 2.8--5.6\% ($f = 2.0$) in these configurations.

The tested Moffat and NEA-inflated configurations keep the relevant $gri$ saturation rates in the few-percent regime. Across these tested PSF models, the Gaussian approximation is sufficient for our forecast-level saturation checks, and 1--5\,s exposures consistently mitigate peak-pixel saturation for bright nearby events.

\section{Saturation Threshold Validation}
\label{sec:sat_validation}

Our scheduler uses a threshold of $N_{\rm sat} = 8 \times 10^4\,\mathrm{e}^-$ (Eq.~\ref{eq:msat}) for the central-pixel saturation check. We assess the corresponding bright limits and band dependence using two complementary approaches: (i) empirical measurements from DP1 commissioning data with LSSTComCam, and (ii) end-to-end \texttt{SNANA} simulations that reproduce our scheduler's saturation logic.

\paragraph{Empirical saturation limits from DP1.}
We analyzed $\sim$15,000 WFD-like visits (28--32~s exposures, airmass $\le 2.0$) from Rubin Data Preview~1 \citep[DP1;][]{Rubin_DP1_dataset_2025,RubinDP1_2026} using LSSTComCam (9 CCDs). For each visit-detector combination, we extracted point sources (extendedness $< 0.2$, SNR $> 50$) and binned them by PSF magnitude in 0.1~mag steps. {For a magnitude bin centered on $m$, we define
\begin{equation}
f_{\rm sat}(m)=\frac{N_{\rm sat}(m)}{N_{\rm src}(m)},
\end{equation}
where $N_{\rm src}(m)$ is the number of selected point sources in that bin and $N_{\rm sat}(m)$ is the subset with the \texttt{pixelFlags\_saturatedCenter} flag set in the DP1 source catalog. We estimate $m_p$ for $p=\{10,50,90\}$ as the interpolated catalog PSF magnitude at which $f_{\rm sat}(m_p)=p/100$ for that visit-detector combination. Thus $m_{50}$ is an empirical half-saturation crossing in measured PSF magnitude, not a limiting magnitude or a detection-completeness threshold.}

Table~\ref{tab:dp1_saturation} summarizes the common finite-threshold subset for the $gri$ bands, which provide the cleanest statistics for a combined threshold table. To translate the empirical $m_{50}$ crossings onto an SN~Ia redshift scale, we use the public DES-SN5YR peak magnitudes only as a magnitude--redshift check; the Rubin saturation threshold itself comes from the DP1 measurements and the scheduler's Rubin electron-count model, not from DES saturation behavior. We derive these redshift equivalents from the public \texttt{DES-SN5YR\_HD+MetaData.csv} sample and the released \texttt{SALT3.DES5YR} model \citep{Sanchez2024_DES5YRrelease}\footnote{Release documentation: \url{https://des-sn-dr.readthedocs.io/en/latest/4_DISTANCES_COVMAT.html}; released SALT3 model documentation: \url{https://des-sn-dr.readthedocs.io/en/latest/2_LCFIT_MODEL.html}; tagged public release used here: \url{https://github.com/des-science/DES-SN5YR/tree/1.3}.} by evaluating peak observer-frame DES $gri$ magnitudes for the low-redshift subset ($z_{\rm HD}<0.12$), constructing a monotonic magnitude--redshift relation in each band, and inverting it at the DP1 $m_{50}$ values. The median 30~s crossings are $m_{50}\approx16.6$--17.0~mag in $gri$; a normal SN~Ia near peak therefore approaches this empirical regime at approximately $z\sim0.045$--0.048 in $g/r$ and $z\sim0.027$--0.030 in $i$, with lower-redshift events being brighter and therefore more susceptible to saturation. The median catalog transition width, $|m_{90}-m_{10}|$, is only 0.14--0.16~mag. We do not assign physical redshift equivalents to $m_{10}$ or $m_{90}$: once a source saturates, its catalog PSF flux can be biased, and in 93--97\% of the common finite-threshold visit-detector subsets the measured ordering is $m_{10}<m_{50}<m_{90}$, opposite to the ordering expected for unbiased source magnitudes. The $m_{50}$ comparison is therefore a consistency check rather than a precision calibration of the full saturation curve. It is consistent with the $8 \times 10^4\,\mathrm{e}^-$ threshold adopted in our scheduler when accounting for typical Rubin zeropoints and exposure times.

\begin{table*}[t]
\centering
\small
\setlength{\tabcolsep}{5.0pt}
\caption{DP1 Empirical Saturation Thresholds for 30~s WFD-like Exposures}
\label{tab:dp1_saturation}
\begin{tabular}{lcccc}
\hline
Band & $N_{\rm valid}$ & $m_{50}$ (mag) & $z_{\rm typ}(m_{50})$ & $|m_{90}-m_{10}|$ (mag) \\
\hline
$g$ & 515 & 16.95 & 0.047 & 0.155 \\
$r$ & 739 & 17.00 & 0.045 & 0.158 \\
$i$ & 673 & 16.60 & 0.027 & 0.138 \\
\hline
\end{tabular}
\tablecomments{Median thresholds derived from DP1 LSSTComCam data \citep{Rubin_DP1_dataset_2025,RubinDP1_2026}. $N_{\rm valid}$ is the number of visit-detector combinations with finite estimates of $m_{10}$, $m_{50}$, and $m_{90}$, so all reported quantities are summarized over the same subset. The table is restricted to $gri$ because those bands provide the cleanest statistics for a combined threshold summary. $z_{\rm typ}(m_{50})$ places the Rubin-derived $m_{50}$ crossing on an approximate SN~Ia redshift scale using DES-SN5YR peak magnitudes; it is a diagnostic scale rather than a detection boundary. The final column is the median absolute per-visit catalog-magnitude range between the 10\% and 90\% crossings; because saturated-source PSF fluxes can be biased, it is quoted only as a transition-width diagnostic.}
\end{table*}

\paragraph{Band dependence and saturation rates.}

The separate \texttt{SNANA} saturation-validation simulations examine the adopted bright-limit model across the LSST filters. They are distinct from the main light-curve forecast runs described in Section~\ref{sec:snana}.
 For a representative low-$z$ SN~Ia sample ($z < 0.05$) observed with 30~s WFD exposures, the epoch-level saturation rates are 2.72\% in $g$, 3.43\% in $r$, and 2.05\% in $i$, where the rate is the fraction of individual observations flagged as saturated. The $z$ band shows negligible saturation ($0.02\%$) because SN~Ia SEDs are fainter at these observed wavelengths and because the empirical DP1 saturation regime is less relevant there for the nearby-SN sample considered here. These simulations are consistent with both the DP1 thresholds and the adopted scheduler saturation check.

Figure~\ref{fig:saturated_lc_examples} shows three representative light curves from the realistic-PSF WFD validation simulation, with the flagged epochs marked explicitly.
\begin{figure*}[t]
\centering
\includegraphics[width=0.96\textwidth]{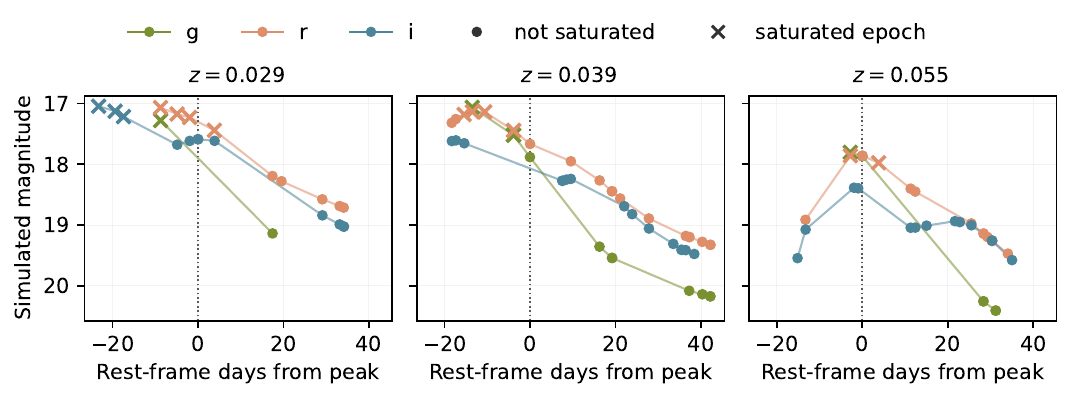}

\caption{Representative low-redshift SN~Ia light curves from the realistic-PSF WFD saturation validation. Lines connect noiseless simulated magnitudes at the visit times; filled circles mark unsaturated observations, and crosses mark central-pixel saturations. Because saturated flux measurements are discarded, crosses are plotted at the corresponding noiseless simulated magnitudes. The vertical dotted line marks simulated peak brightness.}

\label{fig:saturated_lc_examples}
\end{figure*}

\paragraph{Host galaxy flux contribution.}
Local host light contributes to the photometric noise and can also affect saturation. We estimate this contribution for 20 low-$z$ SNe~Ia ($0.003<z<0.06$) placed at the centers of their assigned simulated host galaxies. Within a region defined by a 1\arcsec-FWHM PSF, the median host flux is ${\sim}0.2$ of the total SN peak flux in $g$, ${\sim}0.3$ in $r$, and ${\sim}0.7$ in $i$, with broad event-to-event variation. We compare matched simulations with and without host flux; including the host flux produced no additional saturated epochs in this test.

\paragraph{WFD recovery potential.}
One alternative to short-exposure observations is to attempt recovery of saturated WFD photometry through pixel-masking techniques, in which saturated pixels are excluded from PSF fits. Such approaches can extend dynamic range for moderately saturated sources, but generally increase photometric uncertainties and can degrade color information when multiple bands saturate near peak. For the brightest SNe at $z \lesssim 0.03$, simultaneous $gri$ saturation is possible and would leave only $z$-band information for SALT3 color constraints. The epoch-level rates above do not by themselves establish how often all three bands ($gri$) saturate in the same SN, which should be measured directly in a larger validation sample. Short (1--5~s) exposures are designed to avoid this regime and provide multiband coverage at peak.
}

\section{\texorpdfstring{Twilight $m_5$ versus Airmass and Sky Region}{Twilight m5 versus Airmass and Sky Region}}

\label{app:m5_zenith_sky}

\begin{figure*}

\centering
\includegraphics[width=\textwidth]{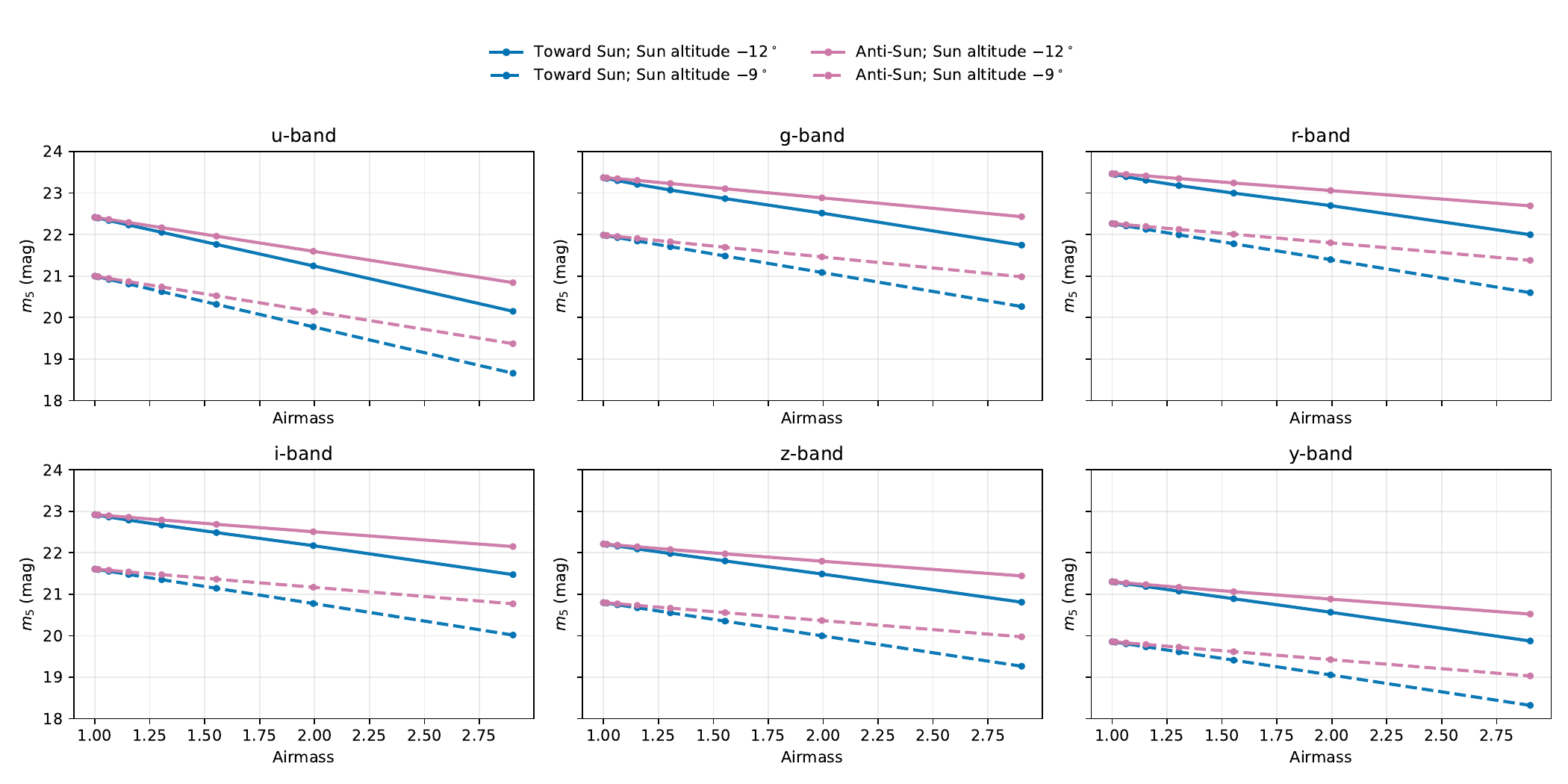}
\caption{\textbf{Twilight 5$\sigma$ depth versus airmass and sky region.} Predicted 30\,s single-visit depth $m_5$ (SNR=5; Equation~\ref{eq:m5}) for Rubin $u/g/r/i/z/y$. Each panel uses the same 18--24\,mag vertical scale. Blue and purple curves give the toward-Sun and anti-Sun sky regions; solid and dashed curves give Sun altitudes of $-12^\circ$ and $-9^\circ$. The points are the original eight airmass-grid evaluations from the RubinSim parametric twilight sky model, connected to guide the eye; scheduled 1--5\,s visit depths are evaluated separately. The $-9^\circ$ curves extrapolate beyond the twilight fit's $-11^\circ$ calibration boundary.}
\label{fig:app_m5_zenith_sky}
\end{figure*}

Figure~\ref{fig:app_m5_zenith_sky} illustrates how predicted twilight depth varies with solar altitude, airmass, and position relative to the Sun. Table~\ref{tab:app_m5_zenith_values} gives the same calculation at selected zenith angles, which correspond to the plotted airmasses.
 We compute $m_5$ with Eq.~\ref{eq:m5} at two representative Sun altitudes that bracket our nominal window, $\mathrm{Alt}_\odot=-12^\circ$ and $-9^\circ$, and for two limiting sky regions parameterized by azimuth relative to the Sun: toward-Sun ($\mathrm{azRelSun}=0^\circ$) and anti-Sun ($180^\circ$). The $-12^\circ$ curves are within the fitted twilight-model domain, whereas the $-9^\circ$ curves are extrapolations and should not be interpreted as calibrated predictions. Scheduled-visit depths are calculated separately at each target position, time, and exposure length. Other exposure times follow Eq.~\ref{eq:m5}, including both the explicit $t_{\rm vis}$ term and the short-exposure correction $\Delta C_m(\tau)$.

Table~\ref{tab:app_m5_zenith_values} gives representative values from the same calculation.

\begin{table*}
\caption{Representative $m_5$ values (SNR=5) for Figure~\ref{fig:app_m5_zenith_sky}\label{tab:app_m5_zenith_values}}
\centering
\small
\setlength{\tabcolsep}{5.0pt}
\renewcommand{\arraystretch}{1.12}
\begin{tabular}{llrccccc}
\hline\hline
Band & Sky region (azRelSun) & $\mathrm{Alt}_\odot$ (deg) & \multicolumn{5}{c}{$m_5$ (mag) at zenith angle $\gamma$ (deg)} \\
\cline{4-8}
 & & & 0 & 30 & 50 & 60 & 70 \\
\hline
u & Toward Sun ($0^\circ$) & -12 & 22.42 & 22.23 & 21.76 & 21.25 & 20.15 \\
u & Toward Sun ($0^\circ$) & -9 & 21.00 & 20.81 & 20.32 & 19.78 & 18.66 \\
u & Anti-Sun ($180^\circ$) & -12 & 22.42 & 22.29 & 21.96 & 21.60 & 20.84 \\
u & Anti-Sun ($180^\circ$) & -9 & 21.00 & 20.87 & 20.52 & 20.15 & 19.37 \\
g & Toward Sun ($0^\circ$) & -12 & 23.37 & 23.21 & 22.87 & 22.52 & 21.75 \\
g & Toward Sun ($0^\circ$) & -9 & 21.99 & 21.84 & 21.48 & 21.09 & 20.26 \\
g & Anti-Sun ($180^\circ$) & -12 & 23.37 & 23.31 & 23.10 & 22.88 & 22.43 \\
g & Anti-Sun ($180^\circ$) & -9 & 21.99 & 21.91 & 21.69 & 21.46 & 20.98 \\
r & Toward Sun ($0^\circ$) & -12 & 23.46 & 23.31 & 23.00 & 22.70 & 22.00 \\
r & Toward Sun ($0^\circ$) & -9 & 22.27 & 22.13 & 21.78 & 21.40 & 20.60 \\
r & Anti-Sun ($180^\circ$) & -12 & 23.46 & 23.41 & 23.24 & 23.06 & 22.69 \\
r & Anti-Sun ($180^\circ$) & -9 & 22.27 & 22.20 & 22.01 & 21.80 & 21.38 \\
i & Toward Sun ($0^\circ$) & -12 & 22.92 & 22.79 & 22.49 & 22.17 & 21.47 \\
i & Toward Sun ($0^\circ$) & -9 & 21.61 & 21.48 & 21.14 & 20.78 & 20.02 \\
i & Anti-Sun ($180^\circ$) & -12 & 22.92 & 22.85 & 22.68 & 22.51 & 22.15 \\
i & Anti-Sun ($180^\circ$) & -9 & 21.61 & 21.54 & 21.36 & 21.17 & 20.77 \\
z & Toward Sun ($0^\circ$) & -12 & 22.21 & 22.09 & 21.80 & 21.49 & 20.81 \\
z & Toward Sun ($0^\circ$) & -9 & 20.80 & 20.67 & 20.35 & 20.00 & 19.26 \\
z & Anti-Sun ($180^\circ$) & -12 & 22.21 & 22.14 & 21.97 & 21.80 & 21.44 \\
z & Anti-Sun ($180^\circ$) & -9 & 20.80 & 20.73 & 20.56 & 20.37 & 19.97 \\
y & Toward Sun ($0^\circ$) & -12 & 21.30 & 21.18 & 20.89 & 20.57 & 19.87 \\
y & Toward Sun ($0^\circ$) & -9 & 19.86 & 19.73 & 19.41 & 19.06 & 18.32 \\
y & Anti-Sun ($180^\circ$) & -12 & 21.30 & 21.23 & 21.06 & 20.88 & 20.52 \\
y & Anti-Sun ($180^\circ$) & -9 & 19.86 & 19.79 & 19.62 & 19.42 & 19.03 \\
\hline
\end{tabular}
\tablecomments{Depths are computed with Eq.~\ref{eq:m5} using the RubinSim sky-brightness model in its parametric mode (no explicit RA/Dec/MJD) at $t_{\rm vis}=30$\,s; other exposure times are recomputed with the same equation, including $\Delta C_m(\tau)$.}
\end{table*}

\end{document}